\documentclass[aps,prd,reprint,showpacs]{revtex4-1}
\usepackage{amssymb}
\usepackage{amsmath}
\usepackage{graphicx}

\begin{document}

\title{Analytical study of the parametric instability of an oscillating scalar field\\
			 in an expanding universe}
\author{Vladimir A. Koutvitsky}
\author{Eugene M. Maslov}
\email{zheka@izmiran.ru}
\affiliation{Pushkov Institute of Terrestrial Magnetism, Ionosphere and Radio Wave 
Propagation (IZMIRAN) of the Russian Academy of Sciences,\\ Kaluzhskoe Hwy 4, Moscow, Troitsk, Russian Federation, 108840 }
\date{\today }

\begin{abstract}
We investigate the dynamics of the perturbations of the inflaton scalar
field oscillating around a minimum of its effective potential in an
expanding universe. With the assumption of smallness of the ratio of the
Hubble parameter to the oscillation frequency we apply the technique of
separation of fast and slow motions. Considering the oscillation phase and
the energy density as fast and slow variables we derive the Hill equation
for the fluctuation modes in which the energy density is treated as a slowly
varying parameter. We develop a general perturbative approach to solving the
equations of this type, which is based on the Floquet theory and asymptotic
expansions in the vicinity of the solutions with the "frozen" parameters. As
an example, we consider the $\phi ^{2}-\phi ^{4}$ potential and construct
the approximate solutions of the corresponding Lam\'{e} equation. The
obtained solutions are found to be in a good agreement with the
results of the direct numerical integration.
\end{abstract}

\maketitle

\section{INTRODUCTION}

The study of dynamics of the oscillating scalar field in an expanding
space-time is of great importance for the modern cosmology. According to the
standard inflationary scenario \cite{Linde3}, soon after a slow-roll regime
during which the universe expands with a huge velocity, the inflaton scalar
field enters the phase of rapid oscillations around a minimum of the
effective potential. This results in reheating of the universe by a
three-stage process \cite{Kofman}. At the first stage, named preheating, the
classical coherently oscillating inflaton scalar field decays into some
boson species due to parametric resonance. At the second stage these decay
products themselves interact and decay, and, at the third stage, thermalize.

The stage of preheating was studied by many authors, and a lot of
interesting results on parametric resonance have been obtained \cite{Kofman,
Shtan, Boyan, Kofman2, Greene1, Khleb, Kaiser1, Kaiser2, Zlatev, Greene2,
Finkel} (see \cite{Allah, Amin1} for a review). The cosmological expansion
leads to the variation in time of the parameters in the corresponding Hill
equations for the fluctuation modes of both the inflaton field and other
fields interacting with it. In the massless conformal theories these
variations can be eliminated by a conformal transformation, so that in this
case the parametric resonance can be considered in Minkowski space-time \cite%
{Greene1, Kaiser1, Kaiser2}. In massive theories, in the beginning of the
preheating stage, when the Hubble parameter $H$ is of the order of the
inflaton oscillation frequency $\omega $, the cosmological expansion results
in the stochastic-like behaviour of the fluctuations of a scalar field
interacting with the inflaton field, provided that the coupling constant is
not too small \cite{Kofman2}. Notice, that some authors in their studies of
preheating neglect the cosmological expansion at all that permits them to
solve the corresponding Hill equations (with constant parameters) exactly 
\cite{Boyan, Finkel}.

The aim of the present paper is to study the effect of the cosmological
expansion on the parametric resonance in a mathematically consistent way for
small values of $H/\omega \sim \varepsilon $. In this case the rapidly
oscillating system passes slowly through the resonant and nonresonant zones
in the space of parameters, that makes it possible to apply the method of
separation of fast and slow motions and to develop the asymptotic theory of
the parametric resonance. In this paper we concentrate on the study of the
parametric instability of the inflaton field itself,\ leaving aside
interaction with other fields and backreaction. We believe however that our
approach is sufficiently general to be applicable to other equations
describing the parametric resonance at the certain stages of preheating.

The paper is organized as follows. In Sec. II we investigate the coherent
oscillations of the scalar background in the Friedmann-Robertson-Walker
universe. We use the oscillation phase $\theta $ and the energy density $%
\rho $ as fast and slow variables and reduce the field equation to the
system with a fast rotating phase. The averaging of this system over the
oscillation period results in the Van der Pol equations that can be solved
through the quadratures. We demonstrate the smallness of difference between
the solutions of exact and averaged equations. In Sec. III, using the
averaged variables $\theta $ and $\rho $, we derive the Hill equation for
the Fourier modes of the inflaton fluctuations where $\rho $ is considered
as a slowly varying parameter. In Sec. IV we develop a general perturbation
approach to solving equations of this type. Based on the Floquet theory, we
construct the solutions in the form of asymptotic expansions near the known
solutions with "frozen" parameters. To demonstrate the validity of our
approach, in Sec. V we consider the $\phi ^{2}-\phi ^{4}$ inflaton
potential. As is known, in this case the Hill equation for the fluctuation
modes takes the form of the Lam\'{e} equation. At first, assuming the
parameters are "frozen", we find the exact solutions of the Lam\'{e}
equation in all resonant and nonresonant zones (including the borders
between them) with the help of the Lindemann-Stieltjes method. Following our
approach we use these results to construct the solutions on the trajectories
in the space of parameters along which the fluctuation modes evolve due to
expansion of the universe. We show that the solutions obtained are in a good
agreement with the results of direct numerical integration of the Lam\'{e}
equation with a slowly varying parameter. Finally, by analyzing the
contributions of various trajectories, we find in the linear approximation
the shape of the localized field fluctuations, into which the oscillating
background can decay. In Sec. VI some remarks are made concerning the
physical meaning of the results obtained.

\section{EVOLUTION OF THE OSCILLATING SCALAR BACKGROUND IN THE EXPANDING
UNIVERSE}

The homogeneous inflaton scalar field $\phi $ in the flat
Friedmann-Robertson-Walker universe is described by the equation 
\begin{equation}
\phi _{tt}+3H\phi _{t}+V^{\prime }(\phi )=0,  \label{eq1}
\end{equation}%
where $H=a_{t}/a$ is the Hubble parameter, $a(t)$ is the scale factor, $%
V(\phi )$ is the effective potential. We will assume that the universe is
filled only with the field $\phi $ having the effective pressure $\mathrm{p}$
and energy density$\rho $,%
\begin{equation}
\mathrm{p}=\phi _{t}^{2}/2-V(\phi ),\text{\quad }\rho =\phi
_{t}^{2}/2+V(\phi ).  \label{eq4a}
\end{equation}%
This field completely determines evolution of the scale factor $a(t)$
through the Friedmann equation%
\begin{equation}
\left( a_{t}/a\right) ^{2}=\left( 8\pi G/3\right) \rho .  \label{eq3}
\end{equation}%
From Eqs. (\ref{eq1})-(\ref{eq3}) it follows that%
\begin{equation}
\rho _{t}=-3(a_{t}/a)(\mathrm{p}+\rho ),  \label{eq5}
\end{equation}%
\begin{equation}
a_{tt}/a=-\left( 4\pi G/3\right) \left( \rho +3\mathrm{p}\right) .
\label{eq2a}
\end{equation}

After the end of the slow-roll stage, the inflaton field $\phi (t)$ begins
to perform fast damped oscillations around the minimum of the potential $%
V(\phi )$. Equation (\ref{eq1}) describing these oscillations can be
reduced, with allowance for (\ref{eq3}) and (\ref{eq4a}), to the form%
\begin{equation}
\phi _{tt}+2\sqrt{3\pi G}\,[\,\phi _{t}^{2}+2V(\phi )\,]^{1/2}\phi _{t}+V^{\prime
}(\phi )=0,  \label{eq4}
\end{equation}%
and hence can be considered as a dissipative dynamical system written in
terms of the independent variables $\phi $ and $\phi _{t}$. Following ref. 
\cite{Kout1}, we pass from these variables to others, $\theta $ and $\rho $,
using the technique of separating fast and slow motions (see, e.g., \cite%
{Mois}). We set%
\begin{eqnarray}
\phi &=&\varphi (\theta ,\rho ),  \label{eq6} \\
\phi _{t} &=&\omega (\rho )\varphi _{\theta }(\theta ,\rho ),  \label{eq7}
\end{eqnarray}%
where $\varphi (\theta ,\rho )$ is a $2\pi $-periodic solution of the
equation%
\begin{equation}
\omega ^{2}(\rho )\varphi _{\theta \theta }+V^{\prime }(\varphi )=0.
\label{eq8}
\end{equation}%
From Eqs. (\ref{eq4a}), (\ref{eq6}), and (\ref{eq7}) it follows that the
first integral of this equation is the energy density $\rho $,%
\begin{equation}
\frac{1}{2}\omega ^{2}(\rho )\varphi _{\theta }^{2}+V(\varphi )=\rho ,
\label{eq9}
\end{equation}%
and, hence,%
\begin{equation}
\omega ^{-1}(\rho )=\frac{1}{\pi \sqrt{2}}\int_{\varphi _{\min }(\rho
)}^{\varphi _{\max }(\rho )}\frac{d\varphi }{\sqrt{\rho -V(\varphi )}},
\label{eq10}
\end{equation}%
where $V(\varphi _{\min ,\;\max })=\rho $. Equations (\ref{eq6})-(\ref{eq10}%
) completely determine the transformation $\left( \phi ,\phi _{t}\right)
\rightarrow \left( \theta ,\rho \right) $.

In order for the above procedure to be self-consistent, Eqs. (\ref{eq6}) and
(\ref{eq7}) must be supplemented by the compatibility condition%
\begin{equation}
\varphi _{\theta }\theta _{t}+\varphi _{\rho }\rho _{t}=\omega \varphi
_{\theta }.  \label{eq11}
\end{equation}%
The equations determining\ evolution of the variables $\rho $ and $\theta $
are obtained from Eqs. (\ref{eq5}) and (\ref{eq11}) using Eqs. (\ref{eq4a}),
(\ref{eq3}), and (\ref{eq7}):%
\begin{eqnarray}
\rho _{t} &=&-2\sqrt{6\pi G}\rho ^{1/2}\omega ^{2}(\rho )\varphi _{\theta
}^{2},  \label{eq12} \\
\theta _{t} &=&\omega (\rho )+2\sqrt{6\pi G}\rho ^{1/2}\omega ^{2}(\rho
)\varphi _{\rho }\varphi _{\theta }.  \label{eq13}
\end{eqnarray}%
It should be emphasized that these equations are exact and fully equivalent
to Eq. (\ref{eq4}). Similar, but more cumbersome equations can be obtained
for the polar coordinates of the variables $\phi $, $\phi _{t}$ \cite{Ren}.

Integrating $\mathrm{p}+\rho =\omega ^{2}(\rho )\varphi _{\theta }^{2}$ over 
$\theta $ from $0$ to $2\pi $ and denoting $\mathrm{\bar{p}}=\left( 2\pi
\right) ^{-1}\int_{0}^{2\pi }\mathrm{p}(\theta ,\rho )\,d\theta $, we
immediately obtain the Turner's formula for the adiabatic index \cite{Tur}:%
\begin{equation}
\gamma (\rho )=1+\frac{\mathrm{\bar{p}}(\rho )}{\rho }=\frac{\sqrt{2}\omega
(\rho )}{\pi \rho }\int_{\varphi _{\min }(\rho )}^{\varphi _{\max }(\rho )}%
\sqrt{\rho -V(\varphi )}\,d\varphi .  \label{eq14}
\end{equation}%
This formula can also be obtained with the help of the action-angle
variables \cite{Masso}. From Eq. (\ref{eq2a}) it is seen that the adiabatic
index determines the dynamics of the cosmological expansion: the universe
will expand with acceleration if $\gamma <2/3$ and with deceleration if $%
\gamma >2/3$. It follows from Eqs. (\ref{eq10}) and (\ref{eq14}) that
dependence $\gamma (\rho )$ is completely determined by the shape of the
effective potential. Thus, in ref. \cite{Dam}\ it was shown that the
accelerated expansion will continue for a time at the oscillation stage too
if the potential is non-convex in regions not too far from the minimum.

The set of equations (\ref{eq12}), (\ref{eq13}) refers to the class of
systems with a rotating phase. If the field oscillations are fast enough
compared to the velocity of the cosmological expansion, i.e., $H/\omega \sim
\varepsilon \ll 1$, then a generalized averaging method can be used to
simplify the system. In general, the method consists in passing from $\theta
,\rho $ to the new, "averaged", variables $\bar{\theta},\bar{\rho}$ with the
help of asymptotic power series in $\varepsilon $ \cite{Mois, Nayf}. In the
lowest order, we have $\rho =\bar{\rho}+O(\varepsilon ),\;\theta =\bar{\theta%
}+O(\varepsilon )$, and evolution equations for $\bar{\rho}$ and $\bar{\theta%
}$ are derived immediately by averaging over the period $2\pi $ of the
right-hand sides of Eqs. (\ref{eq12}) and (\ref{eq13}). This corresponds to
the Van der Pol approximation. As a result, we find:%
\begin{eqnarray}
\rho _{t} &=&-2\sqrt{6\pi G}\rho ^{3/2}\gamma (\rho ),  \label{eq15} \\
\theta _{t} &=&\omega (\rho ),  \label{eq16}
\end{eqnarray}%
where we have dropped the bars and neglected the term of the order of $%
\varepsilon $ arising from averaging of the second term in Eq. (\ref{eq13}).
The latter is caused by\ neglecting the next order term ($\sim \varepsilon
^{2}$) in Eq. (\ref{eq15}), which leads to an error in $\rho $ of the order
of $\varepsilon ^{2}t$ and, therefore, gives the error in $\omega $ of the
order of $\varepsilon $ for large $t\sim \varepsilon ^{-1}$. This situation
is typical when considering systems with a rotating phase in the lowest
order approximation (see, e.g., \cite{Mois}).

\begin{figure}[htb]
\includegraphics[width=0.48\textwidth]{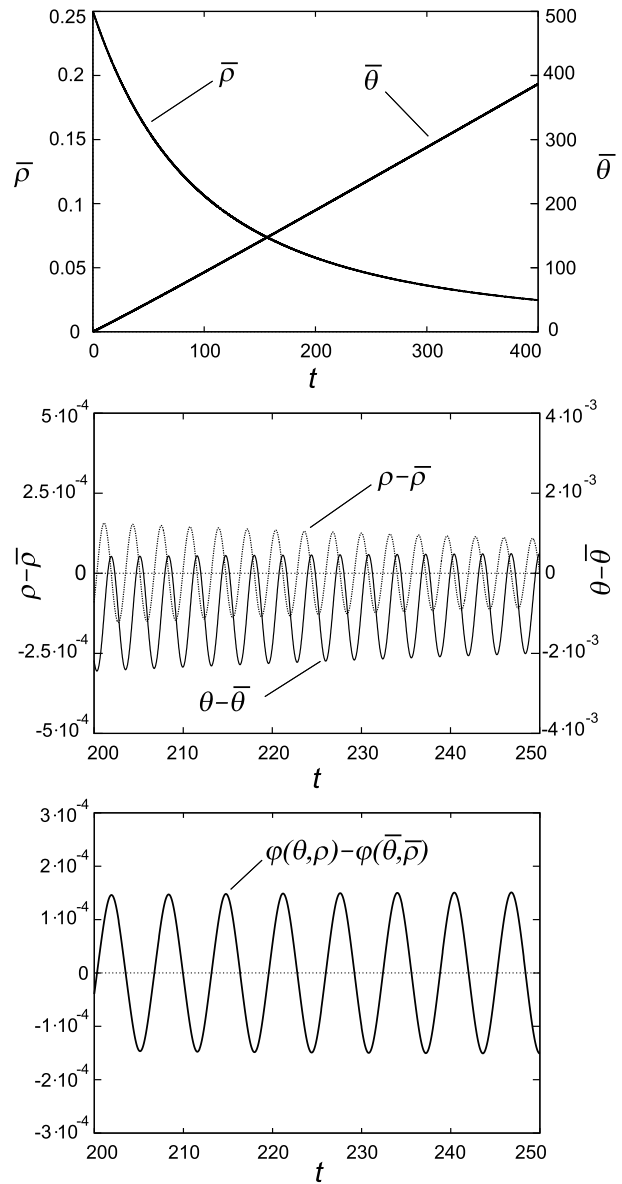}
\caption{The time dependence of the variables $\bar{\protect\theta}$, $\bar{%
\protect\rho}$, and the differences $\protect\theta-\bar{\protect\theta}$, $%
\protect\rho-\bar{\protect\rho}$, and $\protect\varphi(\protect\theta,%
\protect\rho)-\protect\varphi(\bar{\protect\theta}, \bar{\protect\rho})$ for
the $\protect\phi^2-\protect\phi^4$ potential with $m^2=1,\ \protect\lambda%
=0.5,\ \protect\varepsilon=0.005$.}
\end{figure}

To verify the accuracy of the averaging procedure we solved numerically the
systems (\ref{eq12}), (\ref{eq13}) and (\ref{eq15}), (\ref{eq16}) for the
case of the potential $V(\phi )=(m^{2}/2)\phi ^{2}-(\lambda /4)\phi ^{4}$.
It is seen from Fig. 1 that the differences between solutions of these
systems, as well as between the corresponding solutions $\varphi (\theta
,\rho )$, are small (of the order of $\varepsilon =\sqrt{(6/\pi)Gm^{2}%
\lambda ^{-1}}\ll 1$) and do not increase with time, as expected. Thus our
following analysis is based on the system (\ref{eq15}), (\ref{eq16}).

\section{INSTABILITY OF THE SPATIALLY UNIFORM OSCILLATIONS}

In the previous section we studied evolution of the homogeneous oscillating
background caused by the cosmological expansion. We now turn to a close
examination of stability of these oscillations. To do this one must proceed,
instead of Eq. (\ref{eq1}), from the full Klein-Gordon equation%
\begin{equation}
\phi _{tt}+3H\phi _{t}-a^{-2}\Delta \phi +V^{\prime }(\phi )=0.  \label{eq17}
\end{equation}

Let us consider small perturbations around spatially uniform oscillations,%
\begin{equation}
\phi (t,\mathbf{r})=\phi (t)+\delta \phi (t,\mathbf{r}),\qquad \left\vert
\delta \phi \right\vert \ll \left\vert \phi \right\vert .  \label{eq18}
\end{equation}%
Setting 
\begin{equation}
\phi (t)=\varphi (\theta ,\rho ),  \label{eq19}
\end{equation}%
\begin{equation}
\delta \phi (t,\mathbf{r})=a^{-3/2}(\rho )\,\omega ^{-1/2}(\rho )\int
Y(\theta ,\mathbf{k})e^{i\mathbf{kr}}d\mathbf{k,}  \label{eq20}
\end{equation}%
and substituting this into Eq. (\ref{eq17})
in the linear approximation we arrive at the equation%
\begin{equation}
\frac{d^{2}Y}{d\theta ^{2}}+\frac{1}{\omega ^{2}(\rho )}\left[ \left( \frac{k%
}{a(\rho )}\right) ^{2}+V^{\prime \prime }(\varphi (\theta ,\rho ))\right]
Y=0,  \label{eq21}
\end{equation}%
where the terms of the second order in $\varepsilon \sim H/\omega $ were
neglected. The dependence $a(\rho )$ and the slow evolution of $\rho (\theta
)$ are determined by the equations%
\begin{eqnarray}
a_{\rho }/a &=&-\left( 3\rho \gamma (\rho )\right) ^{-1},  \label{eq22} \\
\rho _{\theta }/\rho  &=&-3\left( H/\omega \right) \gamma (\rho ),
\label{eq23}
\end{eqnarray}%
resulting from Eqs. (\ref{eq3}), (\ref{eq15}), and (\ref{eq16}); $H(\rho )$, 
$\omega (\rho )$, and $\gamma (\rho )$ are given by Eqs. (\ref{eq3}), (\ref%
{eq10}), and (\ref{eq14}).

Since $\varphi (\theta ,\rho )$ is a $2\pi $-periodic function in $\theta $,
and $\rho _{\theta }\sim \varepsilon $, equation (\ref{eq21}) belongs to the
class of the Hill equations with a slowly varying parameter. Such equations
describe parametric instability in various physical systems being under the
action of slowly varying factors. We will try to find a general approach to
solving equations of this type. The previous approaches to solving such
equations were based on the assumption that the oscillating terms in the
square brackets in (\ref{eq21}) are small, of the order of $\varepsilon $
(see, e.g., \cite{Burd, Naif1, Naif2, Ng} and references therein). In the
present paper we do not impose this restriction, however, the knowledge of
solutions of the considered equations with $\varepsilon =0$ is assumed.

\section{THE HILL\ EQUATION\ WITH\ A\ SLOWLY\ VARYING\ PARAMETER. GENERAL\
CONSIDERATION}

Let us consider the Hill equation of a general form%
\begin{equation}
d^{2}Y/d\tau ^{2}+h(\tau ,\varkappa )Y=0,  \label{eq24}
\end{equation}%
where $h(\tau ,\varkappa )$ is a $2\pi $-periodic even function with respect
to the fast variable $\tau $, and the parameter $\varkappa $ varies slowly
in accordance with an equation%
\begin{equation}
d\varkappa /d\tau =\varepsilon \,R(\varkappa )\quad (\varepsilon \ll 1).
\label{eq25}
\end{equation}%
In the context of Eq. (\ref{eq21}) the dimensionless variables $\tau $ and $%
\varkappa $ are associated with the phase $\theta $ and the energy density $%
\rho $.

The Floquet theory suggests that the independent solutions of Eq. (\ref{eq24}%
) can be constructed in the form%
\begin{equation}
Y^{\pm }(\tau )=\psi ^{\pm }(\tau ,\varkappa )\,e\,^{\pm \zeta ^{\pm }(\tau
)},  \label{eq26}
\end{equation}%
where $\psi ^{\pm }(\tau ,\varkappa )$ are $2\pi $-periodic or $2\pi $%
-antiperiodic functions in $\tau $, and $d\zeta ^{\pm }/d\tau $ are
effective Floquet exponents. We will seek them through the asymptotic\
expansions%
\begin{eqnarray}
\psi ^{\pm }(\tau ,\varkappa ) &=&\sum_{n=0}^{\infty }\varepsilon ^{n}\psi
_{n}^{\pm }(\tau ,\varkappa ),  \label{eq27} \\
d\zeta ^{\pm }/d\tau &=&\sum_{n=0}^{\infty }\varepsilon ^{n}\mu _{n}^{\pm
}(\varkappa ),  \label{eq28}
\end{eqnarray}%
where $\psi _{0}^{\pm }(\tau ,\varkappa )=\psi _{0}(\pm \tau ,\varkappa ),$ $%
\mu _{0}^{\pm }(\varkappa )=\mu _{0}(\varkappa )$ with normalizing $\psi
_{0}(0,\varkappa )=1,\ \psi _{n}^{\pm }(0,\varkappa )=0$\ $(n\geqslant 1)$, $%
\zeta ^{\pm }(0)=0$.

Substitution of $Y^{\pm }(\tau )$ into Eq. (\ref{eq24}) gives

\begin{equation}
\left[ (\partial /\partial \tau \pm \mu _{0}(\varkappa ))^{2}+h(\tau
,\varkappa )\right] \psi _{n}^{\pm }(\tau ,\varkappa )=F_{n}^{\pm }(\tau
,\varkappa ),  \label{eq29}
\end{equation}%
where $F_{0}^{\pm }=0$, and the functions $F_{n}^{\pm }(\tau ,\varkappa )$
with $n\geqslant 1$, involving $\psi _{m}^{\pm }(\tau ,\varkappa )$ with $m<n
$, are $2\pi $-(anti)periodic in $\tau $:%
\begin{widetext} 
\begin{eqnarray}
F_{1}^{\pm }(\tau ,\varkappa ) &=&-\left( 2\mu _{0}\mu _{1}^{\pm }\pm R\frac{%
d\mu _{0}}{d\varkappa }\right) \psi _{0}^{\pm }\mp 2\mu _{1}^{\pm }\frac{%
\partial \psi _{0}^{\pm }}{\partial \tau }\mp 2R\mu _{0}\frac{\partial \psi
_{0}^{\pm }}{\partial \varkappa }-2R\frac{\partial ^{2}\psi _{0}^{\pm }}{%
\partial \tau \partial \varkappa },  \label{eq30} \\
F_{2}^{\pm }(\tau ,\varkappa ) &=&-\left( \mu _{1}^{\pm 2}+2\mu _{0}\mu
_{2}^{\pm }\pm R\frac{d\mu _{1}^{\pm }}{d\varkappa }\right) \psi _{0}^{\pm
}\mp 2\mu _{2}^{\pm }\frac{\partial \psi _{0}^{\pm }}{\partial \tau }%
-R\left( \frac{dR}{d\varkappa }\pm 2\mu _{1}^{\pm }\right) \frac{\partial
\psi _{0}^{\pm }}{\partial \varkappa }  \notag \\
&&-R^{2}\frac{\partial ^{2}\psi _{0}^{\pm }}{\partial \varkappa ^{2}}-\left(
2\mu _{0}\mu _{1}^{\pm }\pm R\frac{d\mu _{0}}{d\varkappa }\right) \psi
_{1}^{\pm }\mp 2\mu _{1}^{\pm }\frac{\partial \psi _{1}^{\pm }}{\partial
\tau }\mp 2R\mu _{0}\frac{\partial \psi _{1}^{\pm }}{\partial \varkappa }-2R%
\frac{\partial ^{2}\psi _{1}^{\pm }}{\partial \tau \partial \varkappa }%
,...\;.  \label{eq31}
\end{eqnarray}
\end{widetext}Setting $\psi _{n}^{\pm }(\tau ,\varkappa )=X_{n}^{\pm }(\tau
,\varkappa )e^{\mp \mu _{0}(\varkappa )\tau }$, we arrive at the
inhomogeneous Hill equation 
\begin{equation}
\partial ^{2}X_{n}^{\pm }/\partial \tau ^{2}+h(\tau ,\varkappa )X_{n}^{\pm
}=F_{n}^{\pm }(\tau ,\varkappa )\,e\,^{\pm \mu _{0}(\varkappa )\tau },
\label{eq32}
\end{equation}%
where $\varkappa $ is just a parameter. With $n=0$ we have the standard Hill
equation%
\begin{equation}
\partial ^{2}X_{0}^{\pm }/\partial \tau ^{2}+h(\tau ,\varkappa )X_{0}^{\pm
}=0.  \label{eq33}
\end{equation}%
We will suppose that its solutions $X_{0}^{\pm }(\tau ,\varkappa )=\psi
_{0}(\pm \tau ,\varkappa )\,e\,^{\pm \mu _{0}(\varkappa )\tau }$ are known.
Denoting their Wronskian $W\left[ X_{0}^{+},X_{0}^{-}\right] $ as $%
W_{0}(\varkappa )$, we then obtain%
\begin{eqnarray}
&&\underset{n\geqslant 1}{X_{n}^{\pm }(\tau ,\varkappa )}\,=\psi _{n}^{\pm
}(\tau ,\varkappa )\,e^{\pm \mu _{0}\tau }  \notag \\
&&=\frac{1}{W_{0}}\Biggl [X_{0}^{-}(\tau ,\varkappa )\left( \beta _{n}^{\pm
}+\int_{0}^{\tau }\!X_{0}^{+}(\tau ,\varkappa )F_{n}^{\pm }(\tau ,\varkappa
)e^{\pm \mu _{0}\tau }d\tau \right)   \notag \\
&&-X_{0}^{+}(\tau ,\varkappa )\left( \beta _{n}^{\pm }+\int_{0}^{\tau
}\!X_{0}^{-}(\tau ,\varkappa )F_{n}^{\pm }(\tau ,\varkappa )e^{\pm \mu
_{0}\tau }d\tau \right) \Biggr].  \label{eq34}
\end{eqnarray}%

The requirement of periodicity of $\psi _{n}^{\pm }(\tau ,\varkappa )$ in $%
\tau $ leads to the conditions%
\begin{equation}
\beta _{n}^{\pm }(\varkappa )=\frac{1}{e^{\pm 4\pi \mu _{0}}-1}%
\int_{0}^{2\pi }X_{0}^{\pm }(\tau ,\varkappa )F_{n}^{\pm }(\tau ,\varkappa
)e^{\pm \mu _{0}\tau }d\tau ,  \label{eq35}
\end{equation}%
\begin{equation}
\int_{0}^{2\pi }X_{0}^{\mp }(\tau ,\varkappa )F_{n}^{\pm }(\tau ,\varkappa
)e^{\pm \mu _{0}\tau }d\tau =0.  \label{eq36}
\end{equation}%
The orthogonality condition (\ref{eq36}) determines the corrections $\mu
_{n}^{\pm }(\varkappa )$.

Thus, the general solution of Eq. (\ref{eq24}) is written as%
\begin{equation}
Y(\tau )=c^{+}\psi ^{+}(\tau ,\varkappa )e^{\zeta ^{+}(\tau )}+c^{-}\psi
^{-}(\tau ,\varkappa )e^{-\zeta ^{-}(\tau )},  \label{eq37}
\end{equation}%
where $\psi ^{\pm }(\tau ,\varkappa )$ and $\zeta ^{\pm }(\tau )$ are found
from the asymptotic expansions (\ref{eq27}) and (\ref{eq28}). We assume that
at the initial moment of time $\zeta ^{\pm }(0)=0$. The constants $c^{\pm }$
are related to the initial conditions at $\tau =0$ by the formulas%
\begin{equation}
c^{\pm }=\frac{1}{W(\varkappa _{0})}\left[ \frac{1}{2}W^{\mp }(\varkappa
_{0})Y(0)\mp \dot{Y}(0)\right] ,  \label{eq38}
\end{equation}%
\begin{equation}
W(\varkappa )=\frac{1}{2}\left( W^{+}+W^{-}\right) ,\quad W^{\pm }(\varkappa
)=\sum_{n=0}^{\infty }\varepsilon ^{n}W_{n}^{\pm },  \label{eq39}
\end{equation}%
\begin{equation}
W_{n}^{\pm }(\varkappa )=-2\left[ \mu _{n}^{\pm }(\varkappa )\pm \left( 
\frac{\partial \psi _{n}^{\pm }(\tau ,\varkappa )}{\partial \tau }\right)
_{\tau =0}\right] ,  \label{eq40}
\end{equation}%
where $\varkappa _{0}=\varkappa (0)$, and the dot denotes $d/d\tau $. When
deriving Eqs. (\ref{eq38})-(\ref{eq40}) we have taken into account the $2\pi 
$-(anti)periodicity of $\psi ^{\pm }(\tau ,\varkappa )$ in $\tau $ and
normalizing $\psi ^{\pm }(0,\varkappa )=1$.

\subsection{Zero-order approximation}

In the lowest approximation the general solution of Eq. (\ref{eq24}) is
given by%
\begin{eqnarray}
Y(\tau ) &\approx &c^{+}\,\psi _{0}(\tau ,\varkappa (\tau
))\,e\,^{\int_{0}^{\tau }[\mu _{0}(\varkappa )+\varepsilon \mu
_{1}^{+}(\varkappa )]d\tau }  \notag \\
&&+c^{-}\,\psi _{0}(-\tau ,\varkappa (\tau ))\,e^{-\int_{0}^{\tau }[\mu
_{0}(\varkappa )+\varepsilon \mu _{1}^{-}(\varkappa )]d\tau },  \label{eq41}
\end{eqnarray}%
where the periodic function $\psi _{0}(\pm \tau ,\varkappa )$ and the
Floquet exponent $\mu _{0}(\varkappa )$ are derived from the known solutions 
$X_{0}^{\pm }(\tau ,\varkappa )=\psi _{0}(\pm \tau ,\varkappa )\,e\,^{\pm
\mu _{0}(\varkappa )\tau }$ of the Hill equation (\ref{eq33}) with the
"frozen" parameter $\varkappa $. From Eqs. (\ref{eq38})-(\ref{eq40}) it
follows that%
\begin{equation}
c^{\pm }\approx \frac{Y(0)}{2}\mp \frac{\dot{Y}(0)}{W_{0}(\varkappa _{0})},
\label{eq42}
\end{equation}%
where%
\begin{eqnarray}
W_{0}(\varkappa ) &=&W\left[ X_{0}^{+},X_{0}^{-}\right] =W_{0}^{+}=W_{0}^{-}
\notag \\
&=&-2\left[ \mu _{0}(\varkappa )+\left( \frac{\partial \psi _{0}(\tau
,\varkappa )}{\partial \tau }\right) _{\tau =0}\right] .  \label{eq43}
\end{eqnarray}

Notice that in the considered approximation we take into account in Eq. (\ref%
{eq41}) the first-order corrections $\varepsilon \mu _{1}^{\pm }(\varkappa )$%
, because, due to integration, they lead to effects of the same order as the
dependence $\varkappa (\tau )$ in $\psi _{0}(\pm \tau ,\varkappa )$. To find 
$\mu _{1}^{\pm }(\varkappa )$ we proceed from Eq. (\ref{eq36}) with $%
F_{1}^{\pm }(\tau ,\varkappa )$ given by Eq. (\ref{eq30}). Using the $2\pi $%
-(anti)periodicity of $\psi _{0}(\pm \tau ,\varkappa )$ we rewrite Eq. (\ref%
{eq36}) in the form%
\begin{eqnarray}
&&2\mu _{1}^{\pm }\left( \mu _{0}I_{1}+I_{2}\right)  \notag \\
&=&\mp R\left[ \left( d\mu _{0}/d\varkappa \right) I_{1}+2\mu
_{0}I_{3}+2I_{4}\right] ,  \label{eq44}
\end{eqnarray}%
where%
\begin{eqnarray}
I_{1} &=&\int_{0}^{2\pi }\psi _{0}(\tau ,\varkappa )\psi _{0}(-\tau
,\varkappa )\,d\tau ,  \notag \\
I_{2} &=&\int_{0}^{2\pi }\frac{\partial \psi _{0}(\tau ,\varkappa )}{%
\partial \tau }\psi _{0}(-\tau ,\varkappa )\,d\tau ,  \notag \\
I_{3} &=&\int_{0}^{2\pi }\frac{\partial \psi _{0}(\tau ,\varkappa )}{%
\partial \varkappa }\psi _{0}(-\tau ,\varkappa )\,d\tau ,  \notag \\
I_{4} &=&\int_{0}^{2\pi }\frac{\partial ^{2}\psi _{0}(\tau ,\varkappa )}{%
\partial \tau \partial \varkappa }\psi _{0}(-\tau ,\varkappa )\,d\tau .
\label{eq45}
\end{eqnarray}%
\vspace{2pt}

\noindent{It is easy to verify that}%
\begin{eqnarray}
&&I_{3}=\frac{1}{2}\frac{dI_{1}}{d\varkappa },\quad I_{4}=\frac{1}{2}\frac{dI_{2}}{d\varkappa },  \label{eq46}\\
&&\mu _{0}I_{1}+I_{2}=-\pi W_{0},  \label{eq47}
\end{eqnarray}%
where $W_{0}$ is given by Eq. (\ref{eq43}). As a result, from Eq. (\ref{eq44}%
) we obtain%
\begin{equation}
\mu _{1}^{\pm }=\mp \frac{R(\varkappa )}{4}\frac{d}{d\varkappa }\ln
W_{0}^{2}(\varkappa )  \label{eq48}
\end{equation}%
and, hence,%
\begin{equation}
\varepsilon \int_{0}^{\tau }\mu _{1}^{\pm }d\tau =\pm \ln \sqrt{\frac{%
W_{0}(\varkappa _{0})}{W_{0}(\varkappa )}}.  \label{eq48a}
\end{equation}

\subsection{First-order approximation}

In the first-order approximation from Eqs. (\ref{eq37})-(\ref{eq40}) it
follows that 
\begin{widetext}
\begin{eqnarray}
Y(\tau ) &\approx &c^{+}\,[\psi _{0}(\tau ,\varkappa (\tau ))+\varepsilon
\psi _{1}^{+}(\tau ,\varkappa (\tau ))]\,e\,^{\int_{0}^{\tau }[\mu
_{0}(\varkappa )+\varepsilon \mu _{1}^{+}(\varkappa )+\varepsilon ^{2}\mu
_{2}^{+}(\varkappa )]\,d\tau }  \notag \\
&&+c^{-}\,[\psi _{0}(-\tau ,\varkappa (\tau ))+\varepsilon \psi
_{1}^{-}(\tau ,\varkappa (\tau ))]\,e^{-\int_{0}^{\tau }[\mu
_{0}(\varkappa )+\varepsilon \mu _{1}^{-}(\varkappa )+\varepsilon ^{2}\mu
_{2}^{-}(\varkappa )]\,d\tau },  \label{eq49}
\end{eqnarray}%
\end{widetext}where%
\begin{eqnarray}
c^{\pm } &\approx &\frac{Y(0)}{2}\left[ 1\mp \frac{\varepsilon }{2}\left( 
\frac{W_{1}^{+}-W_{1}^{-}}{W_{0}}\right) _{\varkappa =\varkappa _{0}}\right]
\notag \\
&&\mp \frac{\dot{Y}(0)}{W_{0}(\varkappa _{0})}\left[ 1-\frac{\varepsilon }{2}%
\left( \frac{W_{1}^{+}+W_{1}^{-}}{W_{0}}\right) _{\varkappa =\varkappa _{0}}%
\right] .\qquad  \label{eq50}
\end{eqnarray}%
The functions $\psi _{1}^{\pm }(\tau ,\varkappa )$ result from Eq. (\ref%
{eq34}) with $F_{1}^{\pm }(\tau ,\varkappa )$ given by Eq. (\ref{eq30}) and%
\begin{equation}
\beta _{1}^{\pm }=\mp \mu _{1}^{\pm }-R\frac{\left( d\mu _{0}/d\varkappa
\right) J_{1}+2\mu _{0}J_{3}+2J_{4}}{e^{4\pi \mu _{0}}-1},  \label{eq51}
\end{equation}%
where%
\begin{eqnarray}
J_{1} &=&\int_{0}^{2\pi }\psi _{0}^{2}(\tau ,\varkappa )e^{2\mu _{0}\tau
}d\tau ,  \notag \\
J_{2} &=&\int_{0}^{2\pi }\frac{\partial \psi _{0}(\tau ,\varkappa )}{%
\partial \tau }\psi _{0}(\tau ,\varkappa )e^{2\mu _{0}\tau }d\tau ,  \notag
\\
J_{3} &=&\int_{0}^{2\pi }\frac{\partial \psi _{0}(\tau ,\varkappa )}{%
\partial \varkappa }\psi _{0}(\tau ,\varkappa )e^{2\mu _{0}\tau }d\tau , 
\notag \\
J_{4} &=&\int_{0}^{2\pi }\frac{\partial ^{2}\psi _{0}(\tau ,\varkappa )}{%
\partial \tau \partial \varkappa }\psi _{0}(\tau ,\varkappa )e^{2\mu
_{0}\tau }d\tau .  \label{eq52}
\end{eqnarray}%
When deriving Eq. (\ref{eq51}) from Eq. (\ref{eq35}) we have taken into
account the relation%
\begin{equation}
\mu _{0}J_{1}+J_{2}=\frac{{1}}{{2}}\left( e^{4\pi \mu _{0}}-1\right) .
\label{eq53}
\end{equation}%
Notice that $\beta _{1}^{+}=\beta _{1}^{-}$ because of Eq. (\ref{eq48}).

The formulas for $\mu _{2}^{\pm }$ can be obtained from Eq. (\ref{eq36})
with $F_{2}^{\pm }(\tau ,\varkappa )$ given by Eq. (\ref{eq31}).

\section{EXAMPLE: THE LAM\'{E} EQUATION}

Let us return to Eq. (\ref{eq21}) and consider the $\phi ^{2}-\phi ^{4}$
potential%
\begin{equation}
V(\phi )=\frac{m^{2}}{2}\phi ^{2}-\frac{\lambda }{4}\phi ^{4}  \label{eq54}
\end{equation}%
with $\lambda>0$. The potentials of this form can ensure the continuation of
the accelerated expansion of the universe during scalar field oscillations 
\cite{Dam} and lead to formation of the oscillating scalar lumps at the late
times \cite{Amin2, Amin3}.

For the potential (\ref{eq54}) all quantities describing the spatially
uniform oscillations can be evaluated exactly. From Eqs. (\ref{eq9}), (\ref%
{eq10}), and (\ref{eq14}) we find%
\begin{equation}
\varphi (\theta ,\rho )=\varphi _{\mathrm{max}}\,\,\mathrm{sn}\left( \frac{%
2K(\varkappa )}{\pi }\theta,\;\varkappa \right) ,  \label{eq55}
\end{equation}%
\begin{equation}
\varphi _{\mathrm{max}}=\varphi _{\mathrm{exm}}\sqrt{\frac{2\varkappa ^{2}}{%
1+\varkappa ^{2}}},\qquad\frac{\omega }{m}=\frac{\pi }{2K(\varkappa )\sqrt{%
1+\varkappa ^{2}}},  \label{eq56}
\end{equation}%
\begin{equation}
\gamma =\frac{2}{3\varkappa ^{2}K(\varkappa )}\left[ \left( 1+\varkappa
^{2}\right) E(\varkappa )-\left( 1-\varkappa ^{2}\right) K(\varkappa )\right]
,  \label{eq57}
\end{equation}%
where, instead of $\rho $, the dimensionless parameter $\varkappa $ is
introduced,%
\begin{equation}
\varkappa =\frac{1}{\sqrt{\rho /\rho _{\mathrm{exm}}}}\left( 1-\sqrt{1-\rho
/\rho _{\mathrm{exm}}}\right) ,  \label{eq58}
\end{equation}%
$0<\varkappa <1$, $K(\varkappa )$ and $E(\varkappa )$ are complete elliptic
integrals, $\rho _{\mathrm{exm}}=V(\phi _{\mathrm{exm}})=m^{4}/\left(
4\lambda \right) ,\;\phi _{\mathrm{exm}}\equiv \varphi _{\mathrm{exm}}=m/%
\sqrt{\lambda }$. Substitution of Eqs. (\ref{eq54}), (\ref{eq55}) and (\ref{eq56}) into
Eq. (\ref{eq21}) leads to the equation%
\begin{widetext}
\begin{equation}
\frac{d^{2}Y}{d\tau ^{2}}+\frac{K^{2}(\varkappa )}{\pi ^{2}}\left[ \left(
1+p^{2}\right) \left( 1+\varkappa ^{2}\right) -6\varkappa ^{2}{\rm{sn}}%
^{2}\left( \frac{K(\varkappa )}{\pi }\tau ,\;\varkappa \right) \right] Y=0,
\label{eq59}
\end{equation}%
\end{widetext}
where $\tau =2\theta $, $p=k/(ma)$. This is the Lam\'{e} equation. It has
the form of Eq. (\ref{eq24}), since, as follows from Eqs. (\ref{eq22}) and (%
\ref{eq23}), $p$ depends on $\varkappa $,%
\begin{equation}
p^{2}(\varkappa )=p_{0}^{2}\exp \left( -\frac{{4}}{{3}}\int_{\varkappa
}^{\varkappa _{0}}\frac{1-\varkappa ^{2}}{\varkappa (1+\varkappa ^{2})\gamma
(\varkappa )}\;d\varkappa \right) ,  \label{eq60}
\end{equation}%
and $\varkappa $ varies slowly according to Eq. (\ref{eq25}) with%
\begin{equation}
R(\varkappa )=-\frac{\varkappa ^{2}\sqrt{1+\varkappa ^{2}}}{1-\varkappa ^{2}}%
K(\varkappa )\gamma (\varkappa ),\quad \varepsilon =\sqrt{\frac{6Gm^{2}}{\pi
\lambda }}.  \label{eq61}
\end{equation}%
Note that the Lam\'{e} equation of various forms appears in inflationary
theories with other potentials as well \cite{Greene2, Finkel}.

\subsection{The Lam\'{e} equation with constant parameters}

In accordance with our approach, first of all, we need to know the solutions
of the Lam\'{e} equation with "frozen" parameters. Thus, in this subsection
we will assume that $\varkappa $ and $p$ in Eq. (\ref{eq59}) are constants.
In this case, according to the Floquet theorem, the independent solutions of
Eq. (\ref{eq59}) have the form%
\begin{equation}
X_{0}^{\pm }(\tau )=\psi _{0}(\pm \tau )\,e\,^{\pm \mu _{0}\tau },
\label{eq62}
\end{equation}%
where $\psi _{0}(\tau )$ is a $2\pi $-periodic or $2\pi $-antiperiodic
function, $\mu _{0}(\varkappa ,p)$ is the Floquet exponent. They can be
found by the Lindemann-Stieltjes method \cite{Whitt}. The main idea of the
method is in treatment of the periodic function in the Hill equation as a
new "time" variable in each interval of monotonicity. This function can be
bounded, as in Eq. (\ref{eq59}), or even unbounded \cite{Kout2, Kout3, Kout4}%
. For various forms of the Lam{\'{e}} equation, the Lindemann-Stieltjes
method has already been applied by several authors and exact solutions have
been obtained in various parameter domains \cite{Greene1, Kaiser2, Greene2,
Finkel, Maslov}. We will follow ref. \cite{Maslov}\ where the Lam\'{e}
equation of the form (\ref{eq59}) was considered (see also \cite{Kout3} for
a general treatment).

Let us introduce, instead of $\tau $, the new "time" variable%
\begin{equation}
z=\mathrm{sn}^{2}\left( \frac{K(\varkappa )}{\pi }\tau ,\;\varkappa \right)
\label{eq63}
\end{equation}%
and set $Y(\tau )=y(z)$ in each interval of monotonicity of the $2\pi $%
-periodic function $z(\tau )$. Eq. (\ref{eq59}) then becomes 
\begin{eqnarray}
&&4z(1-z)(1-\varkappa ^{2}z)y^{\prime \prime
}+2[1-2(1+\varkappa^{2})z+3\varkappa ^{2}z^{2}]y^{\prime }  \notag \\
&&+[(1+p^{2})(1+\varkappa ^{2})-6\varkappa ^{2}z]y=0  \label{eq64}
\end{eqnarray}%
(the prime denotes $d/dz$).

Consider now any one interval of monotonicity of $z(\tau )$. Denote as $%
y_{1}(z)$ and $y_{2}(z)$ those two linearly independent solutions of Eq. (%
\ref{eq64}) one of which coincides with $X_{0}^{+}(\tau )$ and another with $%
X_{0}^{-}(\tau )$ on the interval chosen. From Eq. (\ref{eq64}), it follows
that the bilinear combinations $y_{1}^{2}$, $y_{2}^{2}$, and $y_{1}y_{2}$
satisfy the third-order equation%
\begin{eqnarray}
&&2z(1-z)(1-\varkappa ^{2}z)Q^{\prime \prime \prime
}+3[1-2(1+\varkappa^{2})z+3\varkappa ^{2}z^{2}]Q^{\prime \prime }  \notag \\
&&+2[p^{2}(1+\varkappa ^{2})-3\varkappa ^{2}z]Q^{\prime }-6\varkappa ^{2}Q=0.
\label{eq65}
\end{eqnarray}%
One of the linearly independent solutions of this equation is the quadratic
polynomial $Q(z)=(z-z_{1})(z-z_{2})$. Its roots are%
\begin{widetext}
\begin{equation}
z_{1,2}=\frac{1}{6\varkappa ^{2}}\left[ (1+\varkappa ^{2})(3-p^{2})\pm \sqrt{%
3(3-p^{2})(1+p^{2})(1+\varkappa ^{2})^{2}-36\varkappa ^{2}}\right] .
\label{eq66}
\end{equation}%
\end{widetext}
Obviously this solution is analytic in $z$ and periodic in $\tau $ on the
whole $\tau $-axis, as well as $X_{0}^{+}(\tau )X_{0}^{-}(\tau )$. Thus, we
identify%
\begin{equation}
y_{1}y_{2}=\frac{(z-z_{1})(z-z_{2})}{z_{1}z_{2}}.  \label{eq67}
\end{equation}%
On the other hand, using Eq. (\ref{eq64}), we obtain%
\begin{equation}
y_{1}y_{2}^{\prime }-y_{2}y_{1}^{\prime }=\frac{C}{\sqrt{q}},  \label{eq68}
\end{equation}%
where $C$ is a constant, and%
\begin{equation}
q(z)=z(1-z)(1-\varkappa ^{2}z).  \label{eq69}
\end{equation}%
It is easy to verify that%
\begin{equation}
q(z_{1})=q(z_{2})=(p^{2}/3)(1+\varkappa ^{2})z_{1}z_{2},  \label{eq70a}
\end{equation}%
\begin{equation}
z_{j}^{2}q-z^{2}q_{j}=z_{j}z(z_{j}-z)(1-\varkappa ^{2}z_{j}z),  \label{eq70b}
\end{equation}%
where $q_{j}=q(z_{j})$ ($j=1,2$) is denoted. From Eqs. (\ref{eq67}) and (\ref%
{eq68}) we find%
\begin{equation}
2\frac{y_{1,2}^{\prime }}{y_{1,2}}=\frac{1}{z-z_{1}}+\frac{1}{z-z_{2}}\mp 
\frac{Cz_{1}z_{2}}{\sqrt{q}(z-z_{1})(z-z_{2})}.  \label{eq71}
\end{equation}%
Substitution of this solution into the corresponding Riccati equation for $%
y^{\prime }/y$ determines $C^{2}$:%
\begin{equation}
C^{2}=q_{j}\left( \frac{z_{1}-z_{2}}{z_{1}z_{2}}\right) ^{2}.  \label{eq72}
\end{equation}%
The choice of the sign of $C$ is arbitrary: changing the sign corresponds to
interchanging of $y_{1}$ and $y_{2}$. Setting $C=\sqrt{q_{j}}(z_{1}-z_{2})/(z_{1}z_{2})$ for definiteness, we rewrite Eq. (\ref{eq71}) as%
\begin{equation}
2\frac{y_{1,2}^{\prime }}{y_{1,2}}=\frac{1}{z-z_{1}}\left( 1\mp \sqrt{\frac{%
q_{j}}{q}}\right) +\frac{1}{z-z_{2}}\left( 1\pm \sqrt{\frac{q_{j}}{q}}%
\right) ,  \label{eq73}
\end{equation}

From Eqs. (\ref{eq66}), (\ref{eq70a}), and (\ref{eq72}) it follows that $%
C^{2}$ is a real quantity. If $C^{2}>0$ the corresponding solutions will
exponentially grow or decay, and if $C^{2}<0$ the solutions will be bounded.
On the $(\varkappa ^{2},p^{2})$-plane these conditions determine two
resonant zones, referred hereinafter as A and B, and two nonresonant zones,
C and D:

\textit{Zone A}%
\begin{eqnarray}
0 &<&p^{2}<\frac{3\varkappa ^{2}}{1+\varkappa ^{2}},  \notag \\
0 &<&z_{2}<1<\varkappa ^{-2}<z_{1}.  \label{eq74}
\end{eqnarray}

\textit{Zone B}%
\begin{eqnarray}
\frac{3}{1+\varkappa ^{2}} &<&p^{2}<1+\frac{2\sqrt{1-\varkappa
^{2}+\varkappa ^{4}}}{1+\varkappa ^{2}},  \notag \\
0 &<&z_{2}<z_{1}<1.  \label{eq75}
\end{eqnarray}

\textit{Zone C}%
\begin{eqnarray}
\frac{3\varkappa ^{2}}{1+\varkappa ^{2}} &<&p^{2}<\frac{3}{1+\varkappa ^{2}},
\notag \\
1 &<&z_{1}<\varkappa ^{-2},\quad z_{2}<0.  \label{eq76}
\end{eqnarray}

\textit{Zone D}%
\begin{eqnarray}
p^{2} &>&1+\frac{2\sqrt{1-\varkappa ^{2}+\varkappa ^{4}}}{1+\varkappa ^{2}},
\notag \\
z_{1} &=&\bar{z}_{1}+i\tilde{z}_{1},\quad \bar{z}_{1}<1,\quad \tilde{z}%
_{1}>0,\quad z_{2}=z_{1}^{\ast }.  \label{eq77}
\end{eqnarray}%
The resulting stability-instability chart is shown in Fig. 2. 


The inequalities (\ref{eq74})-(\ref{eq77}) should be taken into account when
integrating Eq. (\ref{eq73}). For example, in the zones A and B, the right
hand side of Eq. (\ref{eq73}) has the poles on the integration interval, so
that the corresponding integrals are understood as their principal values.
\begin{figure}[ht]
\includegraphics[width=0.47\textwidth]{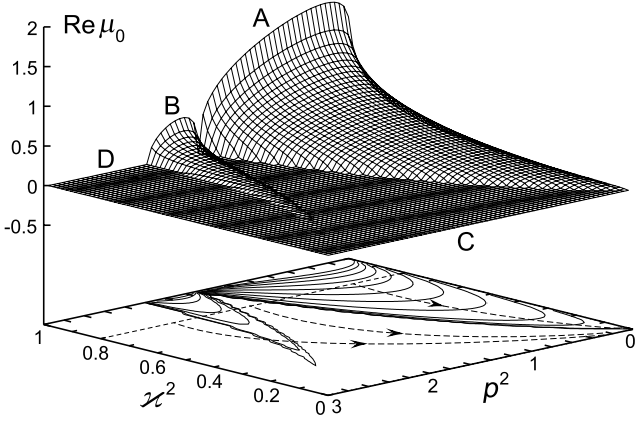}
\caption{The stability-instability landscape for the Lam{\'e} equation (\ref{eq59}).}
\end{figure}
In this case it is convenient to use the identity%
\begin{eqnarray}
\frac{1}{(z-z_{j})\sqrt{q}} &=&-\frac{1}{z_{j}\sqrt{q}}+\frac{1}{%
z_{j}(1-\varkappa ^{2}z_{j}z)\sqrt{q}}  \notag \\
&&+\frac{1}{\sqrt{q_{j}}}\frac{d}{dz}\ln \frac{z_{j}\sqrt{q}-z\sqrt{q_{j}}}{%
z_{j}\sqrt{q}+z\sqrt{q_{j}}}.  \label{eq78}
\end{eqnarray}%
Also, we will use the following representations of the elliptic integrals:%
\begin{eqnarray}
F(\delta ,\varkappa ) &=&\int_{0}^{\delta }\frac{d\vartheta }{\sqrt{%
1-\varkappa ^{2}\sin ^{2}\vartheta }}  \notag \\
&=&\frac{1}{2}\int_{0}^{z}\frac{dz}{\sqrt{q}},  \label{eq79}
\end{eqnarray}%
\begin{eqnarray}
E(\delta ,\varkappa ) &=&\int_{0}^{\delta }\sqrt{1-\varkappa ^{2}\sin
^{2}\vartheta }\,d\vartheta  \notag \\
&=&\frac{1}{2}\int_{0}^{z}\frac{1-\varkappa ^{2}z}{\sqrt{q}}\,dz,
\label{eq80}
\end{eqnarray}%
\begin{eqnarray}
\Pi (\delta ,n,\varkappa ) &=&\int_{0}^{\delta }\frac{d\vartheta }{(1-n\sin
^{2}\vartheta )\sqrt{1-\varkappa ^{2}\sin ^{2}\vartheta }}  \notag \\
&=&\frac{1}{2}\int_{0}^{z}\frac{dz}{(1-nz)\sqrt{q}},  \label{eq81}
\end{eqnarray}%
where $z=\sin ^{2}\delta $.

Taking into account the above, let us construct the solutions (\ref{eq62})
in each of the zones listed and on the borders between them.

\subsubsection{Zone A (resonant)}

In this zone, the integration of Eq. (\ref{eq73}) using the identity (\ref%
{eq78}) gives%
\begin{equation}
y_{1,2}^{2}=c_{1,2}^{2}A_{1,2}^{2}(z,z_{1},z_{2})e^{\pm 2f(z,z_{1},z_{2})},
\label{eq82}
\end{equation}%
where $c_{1,2}^{2}$ are constants,%
\begin{eqnarray}
A_{1}(z,z_{1},z_{2}) &=&\frac{z_{2}-z}{z_{2}\sqrt{q}+z\sqrt{q_{2}}} \\
&&\times\sqrt{z(1-\varkappa ^{2}z_{2}z)(1-z/z_{1})},  \label{eq83} \\
A_{2}(z,z_{1},z_{2}) &=&\frac{z_{2}\sqrt{q}+z\sqrt{q_{2}}}{z_{2}}\sqrt{\frac{%
1-z/z_{1}}{z(1-\varkappa ^{2}z_{2}z)}},  \label{eq84} \\
f(z,z_{1},z_{2}) &=&\frac{\sqrt{q_{1}}}{z_{1}}\Pi
(\delta,z_{1}^{-1},\varkappa )+s(z,z_{2}),  \label{eq85}
\end{eqnarray}%
and we introduce, also for the further,%
\begin{equation}
s(z,z_{j})=\frac{\sqrt{q_{j}}}{z_{j}}\left[ \Pi (\delta ,\varkappa
^{2}z_{j},\varkappa )-F(\delta ,\varkappa )\right] .  \label{eq86}
\end{equation}%
From Eqs. (\ref{eq83}) and (\ref{eq84}) it is clear that $y_{1}$ can become
zero (when $z(\tau )=z_{2}$), and $y_{2}$ can not. This relates with $2\pi $%
-antiperiodicity, and, hence, $4\pi $-periodicity of the function $\psi
_{0}(\tau )$ in the solutions (\ref{eq62}). Thus, to construct the solution $%
X_{0}^{+}=\psi _{0}(\tau )e^{\mu _{0}\tau }$ we set: $X_{0}^{+}(\tau
)=y_{1}(z)\;(0\leqslant \tau \leqslant \pi )$, $X_{0}^{+}(\tau
)=y_{2}(z)\;(\pi \leqslant \tau \leqslant 2\pi )$, $X_{0}^{+}(\tau
)=y_{1}(z)\;(2\pi \leqslant \tau \leqslant 3\pi )$, $X_{0}^{+}(\tau
)=y_{2}(z)\;(3\pi \leqslant \tau \leqslant 4\pi )$. The constants $c_{1,2}$
are different on the different intervals. They are determined by the
matching conditions starting from $X_{0}^{+}(0)=1$. Then $X_{0}^{+}(4\pi
)=e^{4f(1,z_{1},z_{2})}=e^{4\pi \mu _{0}}$, that gives the Floquet exponent%
\begin{equation}
\mu _{0}=\frac{1}{\pi }f(1,z_{1},z_{2}).  \label{eq87}
\end{equation}%
As a result we obtain%
\begin{eqnarray}
\underset{0\leqslant \tau \leqslant \pi }{\psi _{0}(\tau )}
&=&A_{1}(z,z_{1},z_{2})e^{f(z,z_{1},z_{2})-\mu _{0}\tau },  \notag \\
\underset{\pi \leqslant \tau \leqslant 2\pi }{\psi _{0}(\tau )}
&=&-A_{2}(z,z_{1},z_{2})e^{-f(z,z_{1},z_{2})-\mu _{0}(\tau -2\pi )},  \notag
\\
\underset{2\pi \leqslant \tau \leqslant 3\pi }{\psi _{0}(\tau )}
&=&-A_{1}(z,z_{1},z_{2})e^{f(z,z_{1},z_{2})-\mu _{0}(\tau -2\pi )},  \notag
\\
\underset{3\pi \leqslant \tau \leqslant 4\pi }{\psi _{0}(\tau )}
&=&A_{2}(z,z_{1},z_{2})e^{-f(z,z_{1},z_{2})-\mu _{0}(\tau -4\pi )}.
\label{eq88}
\end{eqnarray}%
To find $\psi _{0}(-\tau )$ on the interval $0\leqslant \tau \leqslant \pi $
we take $\psi _{0}(\tau )$ on the interval $3\pi \leqslant \tau \leqslant
4\pi $ and replace $\tau \rightarrow -\tau +4\pi $, for $\psi _{0}(-\tau )$
on the interval $\pi \leqslant \tau \leqslant 2\pi $ we take $\psi _{0}(\tau
)$ on the interval $2\pi \leqslant \tau \leqslant 3\pi $ and replace $\tau
\rightarrow -\tau +4\pi $, and so on. In this way we find%
\begin{eqnarray}
\underset{0\leqslant \tau \leqslant \pi }{\psi _{0}(-\tau )}
&=&A_{2}(z,z_{1},z_{2})e^{-f(z,z_{1},z_{2})+\mu _{0}\tau },  \notag \\
\underset{\pi \leqslant \tau \leqslant 2\pi }{\psi _{0}(-\tau )}
&=&-A_{1}(z,z_{1},z_{2})e^{f(z,z_{1},z_{2})+\mu _{0}(\tau -2\pi )},  \notag
\\
\underset{2\pi \leqslant \tau \leqslant 3\pi }{\psi _{0}(-\tau )}
&=&-A_{2}(z,z_{1},z_{2})e^{-f(z,z_{1},z_{2})+\mu _{0}(\tau -2\pi )},  \notag
\\
\underset{3\pi \leqslant \tau \leqslant 4\pi }{\psi _{0}(-\tau )}
&=&A_{1}(z,z_{1},z_{2})e^{f(z,z_{1},z_{2})+\mu _{0}(\tau -4\pi )}.
\label{eq89}
\end{eqnarray}%
It is seen that $\psi _{0}(\tau )\psi _{0}(-\tau
)=A_{1}A_{2}=(z-z_{1})(z-z_{2})/(z_{1}z_{2})$, in agreement with Eq. (\ref%
{eq67}). Using Eq. (\ref{eq43}) we obtain the Wronskian:%
\begin{equation}
W_{0}=\frac{2}{\pi }\frac{z_{1}-z_{2}}{\sqrt{z_{1}z_{2}}}K(\varkappa )\sqrt{%
(p^{2}/3)(1+\varkappa ^{2})}.  \label{eq90}
\end{equation}

\subsubsection{Zone B (resonant)}

In this zone, integrating Eq. (\ref{eq73}), we obtain%
\begin{eqnarray}
y_{1}^{2} &=&c_{1}^{2}B^{2}(z,z_{2},z_{1})e^{-2g(z,z_{1},z_{2})},  \notag \\
y_{2}^{2} &=&c_{2}^{2}B^{2}(z,z_{1},z_{2})e^{2g(z,z_{1},z_{2})},
\label{eq91}
\end{eqnarray}%
where $c_{1,2}^{2}$ are constants,%
\begin{eqnarray}
B(z,z_{1},z_{2}) &=&\frac{z_{1}-z}{z_{2}}\frac{z_{2}\sqrt{q}+z\sqrt{q_{2}}}{%
z_{1}\sqrt{q}+z\sqrt{q_{1}}}\sqrt{\frac{1-\varkappa ^{2}z_{1}z}{1-\varkappa
^{2}z_{2}z}},  \notag \\
g(z,z_{1},z_{2}) &=&s(z,z_{1})-s(z,z_{2}),  \label{eq92}
\end{eqnarray}%
and $s(z,z_{j})$ is defined by Eq. (\ref{eq86}). It is seen, that $y_{1}$
and $y_{2}$ become zero at $z=z_{2}$ and $z=z_{1}$, respectively. This
ensures the $2\pi $-periodicity of $\psi _{0}(\tau )$ in Eq. (\ref{eq62}).
Thus, we set $X_{0}^{+}(\tau )=y_{2}(z)\;(0\leqslant \tau \leqslant \pi )$, $%
X_{0}^{+}(\tau )=y_{1}(z)\;(\pi \leqslant \tau \leqslant 2\pi )$. Matching $%
y_{2}$ and $y_{1}$, we obtain%
\begin{equation}
\mu _{0}=\frac{1}{\pi }g(1,z_{1},z_{2}),  \label{eq93}
\end{equation}%
\begin{eqnarray}
\underset{0\leqslant \tau \leqslant \pi }{\psi _{0}(\tau )}
&=&B(z,z_{1},z_{2})e^{g(z,z_{1},z_{2})-\mu _{0}\tau },  \notag \\
\underset{\pi \leqslant \tau \leqslant 2\pi }{\psi _{0}(\tau )}
&=&B(z,z_{2},z_{1})e^{-g(z,z_{1},z_{2})-\mu _{0}(\tau -2\pi )}  \label{eq94}
\end{eqnarray}%
and, hence,%
\begin{eqnarray}
\underset{0\leqslant \tau \leqslant \pi }{\psi _{0}(-\tau )}
&=&B(z,z_{2},z_{1})e^{-g(z,z_{1},z_{2})+\mu _{0}\tau },  \notag \\
\underset{\pi \leqslant \tau \leqslant 2\pi }{\psi _{0}(-\tau )}
&=&B(z,z_{1},z_{2})e^{g(z,z_{1},z_{2})+\mu _{0}(\tau -2\pi )}.  \label{eq95}
\end{eqnarray}%
Again we see that $\psi _{0}(\tau )\psi _{0}(-\tau
)=(z-z_{1})(z-z_{2})/(z_{1}z_{2})$. The Wronskian of the obtained solutions
is%
\begin{equation}
W_{0}=-\frac{2}{\pi }\frac{z_{1}-z_{2}}{\sqrt{z_{1}z_{2}}}K(\varkappa )\sqrt{%
(p^{2}/3)(1+\varkappa ^{2})}.  \label{eq96}
\end{equation}

\subsubsection{Zone C (nonresonant)}

Since in this zone $z_{1}>1$, $z_{2}<0$, there are no nonintegrable
singularities in Eq. (\ref{eq73}). This means, that the solutions (\ref{eq62}%
) are nonresonant, with pure imaginary Floquet exponent $\mu _{0}=i\nu _{0}$%
. The corresponding formulas follow immediately from Eqs. (\ref{eq92})-(\ref%
{eq96}) and (\ref{eq70b}), where $\sqrt{z_{2}}=i\sqrt{\left\vert
z_{2}\right\vert }$, $\sqrt{q_{j}}=i\sqrt{\left\vert q_{j}\right\vert }$
and, hence,%
\begin{eqnarray}
B(z,z_{2},z_{1}) &=&B^{\ast }(z,z_{1},z_{2}),  \notag \\
g(z,z_{1},z_{2}) &=&i\left\vert g(z,z_{1},z_{2})\right\vert ,  \label{eq97}
\end{eqnarray}%
\begin{equation}
\nu _{0}=\frac{1}{\pi }\left\vert g(1,z_{1},z_{2})\right\vert .  \label{eq98}
\end{equation}%
It is seen that $\psi _{0}(\tau )$ is $2\pi $-periodic and $\psi _{0}(-\tau
)=\psi _{0}^{\ast }(\tau )$.

\subsubsection{Zone D (nonresonant)}

In this zone $z_{1}$ and $z_{2}$ are complex numbers and $z_{2}=z_{1}^{\ast
} $. The solutions (\ref{eq62}), obtained by integrating Eq. (\ref{eq73}),
are nonresonant, with $\mu _{0}=i\nu _{0}$. The corresponding formulas
follow from Eqs. (\ref{eq92})-(\ref{eq96}), where $z_{2}=z_{1}^{\ast }=\bar{z%
}_{1}-i\tilde{z}_{1}$, $q_{j}>0$,%
\begin{eqnarray}
B(z,z_{1},z_{1}^{\ast }) &=&B^{\ast }(z,z_{1}^{\ast },z_{1})=\frac{z_{1}-z}{%
z_{1}^{\ast }}e^{-i(\alpha +2\beta )},  \notag \\
\tan \alpha &=&\frac{\varkappa ^{2}\tilde{z}_{1}z}{1-\varkappa ^{2}\bar{z}%
_{1}z},\quad \tan \beta =\frac{\tilde{z}_{1}\sqrt{q}}{\bar{z}_{1}\sqrt{q}+z%
\sqrt{q_{1}}},  \notag \\
g(z,z_{1},z_{1}^{\ast }) &=&2i\mathrm{Im}s(z,z_{1})=i\left\vert
g(z,z_{1},z_{1}^{\ast })\right\vert ,  \label{eq99}
\end{eqnarray}%
\begin{equation}
\nu _{0}=\frac{1}{\pi }\left\vert g(1,z_{1},z_{1}^{\ast })\right\vert .
\label{eq100}
\end{equation}%
Again $\psi _{0}(\tau )$ is $2\pi $-periodic and $\psi _{0}(-\tau )=\psi
_{0}^{\ast }(\tau )$.

\subsubsection{Border AC}

To obtain the solutions on the border between zone A and zone C we take the
following linear combinations of the solutions in zone A:%
\begin{eqnarray}
Y_{1} &=&\frac{\sqrt{z_{2}}}{2}\left[ \psi _{0}(\tau )e^{\mu _{0}\xi }-\psi
_{0}(-\tau )e^{-\mu _{0}\xi }\right]  \notag \\
&=&\chi _{1}(\tau )\cosh \mu _{0}\xi +\sqrt{z_{2}}\chi _{2}(\tau )\sinh \mu
_{0}\xi ,  \notag \\
Y_{2} &=&\frac{1}{2}\left[ \psi _{0}(\tau )e^{\mu _{0}\xi }+\psi _{0}(-\tau
)e^{-\mu _{0}\xi }\right]  \notag \\
&=&\chi _{2}(\tau )\cosh \mu _{0}\xi +\frac{1}{\sqrt{z_{2}}}\chi _{1}(\tau
)\sinh \mu _{0}\xi ,  \label{eq101}
\end{eqnarray}%
where%
\begin{eqnarray}
\chi _{1}(\tau ) &=&\frac{\sqrt{z_{2}}}{2}\left[ \psi _{0}(\tau )-\psi
_{0}(-\tau )\right] ,  \notag \\
\chi _{2}(\tau ) &=&\frac{1}{2}\left[ \psi _{0}(\tau )+\psi _{0}(-\tau )%
\right] ,  \label{eq103}
\end{eqnarray}%
$\mu _{0}$ and $\psi _{0}(\pm \tau )$ are given by Eqs. (\ref{eq87})-(\ref%
{eq89}), $\xi =\tau -\tau _{AC}$, $\tau _{AC}$ is some constant. On the
border AC 
\begin{equation*}
p^{2}=3\varkappa ^{2}/(1+\varkappa ^{2}),\quad z_{1}=\varkappa^{-2},\quad
z_{2}=0.
\end{equation*}
Hence, as follows from the identity (\ref{eq70a}), in the vicinity of the
border%
\begin{equation}
z_{1}\approx \frac{1}{\varkappa ^{2}}\left( 1+\frac{\varkappa ^{4}}{%
1-\varkappa ^{2}}\Delta \right) ,\quad z_{2}\approx \Delta ,  \label{eq104}
\end{equation}%
where $\Delta =q(z_{1})=q(z_{2})\ll 1$. When approaching the border from the
side of zone A we have (see Eqs. (\ref{eq83})-(\ref{eq87}) when $\Delta
\rightarrow 0$)%
\begin{eqnarray}
A_{1,2}(z,z_{1},z_{2}) &\rightarrow &\mp \sqrt{z(1-\varkappa ^{2}z)}\Delta
^{-1/2}+a(z,\varkappa ),  \notag \\
f(z,z_{1},z_{2}) &\rightarrow &u(z,\varkappa )\Delta ^{1/2},  \notag \\
\mu _{0} &\rightarrow &\frac{u(1,\varkappa )}{\pi }\Delta ^{1/2},
\label{eq105}
\end{eqnarray}%
where%
\begin{eqnarray}
a(z,\varkappa ) &=&(1-\varkappa ^{2}z)\sqrt{1-z},  \notag \\
u(z,\varkappa ) &=&\varkappa ^{2}\Pi (\delta ,\varkappa ^{2},\varkappa
)+F(\delta ,\varkappa )-E(\delta ,\varkappa ).  \label{eq108}
\end{eqnarray}%
As a result, from Eqs. (\ref{eq88}), (\ref{eq89}), (\ref{eq101}), and (\ref%
{eq103}) we obtain the following solutions on the border AC:%
\begin{eqnarray}
Y_{1} &=&\chi _{1}(\tau ),  \notag \\
Y_{2} &=&\chi _{2}(\tau )+u(1,\varkappa )\chi _{1}(\tau )\xi /\pi ,
\label{eq109}
\end{eqnarray}%
where $2\pi $-antiperiodic functions $\chi _{1,2}(\tau )$ are given by%
\begin{eqnarray}
&&\underset{0\leqslant \tau \leqslant 2\pi }{\chi _{1}(\tau )} =-\sqrt{%
z(1-\varkappa ^{2}z)},  \notag \\
&&\underset{2\pi \leqslant \tau \leqslant 4\pi }{\chi _{1}(\tau )}=\sqrt{%
z(1-\varkappa ^{2}z)},  \notag \\
&&\underset{0\leqslant \tau \leqslant \pi }{\chi _{2}(\tau )} =\phantom{-}%
a(z,\varkappa)  \notag \\
&&-\sqrt{z(1-\varkappa ^{2}z)}\left[ u(z,\varkappa )-u(1,\varkappa )\tau
/\pi \right] ,  \notag \\
&&\underset{\pi \leqslant \tau \leqslant 2\pi }{\chi _{2}(\tau )}
=-a(z,\varkappa )  \notag \\
&&+\sqrt{z(1-\varkappa ^{2}z)}\left[ u(z,\varkappa)+u(1,\varkappa )(\tau
-2\pi )/\pi \right] ,  \notag \\
&&\underset{2\pi \leqslant \tau \leqslant 3\pi }{\chi _{2}(\tau )}%
=-a(z,\varkappa )  \notag \\
&&+\sqrt{z(1-\varkappa ^{2}z)}\left[ u(z,\varkappa)-u(1,\varkappa )(\tau
-2\pi )/\pi \right] ,  \notag \\
&&\underset{3\pi \leqslant \tau \leqslant 4\pi }{\chi _{2}(\tau )}=%
\phantom{-}a(z,\varkappa )  \notag \\
&&-\sqrt{z(1-\varkappa ^{2}z)}\left[ u(z,\varkappa)+u(1,\varkappa )(\tau
-4\pi )/\pi \right].\quad  \label{eq110}
\end{eqnarray}

\subsubsection{Border BC}

In order to find solutions on the borders of zone B we consider the
following linear combinations:%
\begin{eqnarray}
Y_{1} &=&\frac{\sqrt{z_{2}/z_{1}}}{2(1-z_{2}/z_{1})}\left[ \psi _{0}(\tau
)e^{\mu _{0}\xi }-\psi _{0}(-\tau )e^{-\mu _{0}\xi }\right]  \notag \\
&=&\chi _{1}(\tau )\cosh \mu _{0}\xi +\frac{\sqrt{z_{2}/z_{1}}}{1-z_{2}/z_{1}%
}\chi _{2}(\tau )\sinh \mu _{0}\xi ,  \notag \\
Y_{2} &=&\frac{1}{2}\left[ \psi _{0}(\tau )e^{\mu _{0}\xi }+\psi _{0}(-\tau
)e^{-\mu _{0}\xi }\right]  \notag \\
&=&\chi _{2}(\tau )\cosh \mu _{0}\xi +\frac{1-z_{2}/z_{1}}{\sqrt{z_{2}/z_{1}}%
}\chi _{1}(\tau )\sinh \mu _{0}\xi ,\qquad  \label{eq112}
\end{eqnarray}%
where%
\begin{eqnarray}
\chi _{1}(\tau ) &=&\frac{\sqrt{z_{2}/z_{1}}}{2(1-z_{2}/z_{1})}\left[ \psi
_{0}(\tau )-\psi _{0}(-\tau )\right] ,  \notag \\
\chi _{2}(\tau ) &=&\frac{1}{2}\left[ \psi _{0}(\tau )+\psi _{0}(-\tau )%
\right] ,  \label{eq114}
\end{eqnarray}%
$\mu _{0}$ and $\psi _{0}(\tau )$ are given by Eqs. (\ref{eq93})-(\ref{eq95}%
), $\xi =\tau -\tau _{B}$, $\tau _{B}$ is some constant. On the border BC 
\begin{equation*}
p^{2}=3/(1+\varkappa ^{2}),\quad z_{1}=1,\quad z_{2}=0.
\end{equation*}
In the vicinity of the border%
\begin{equation}
z_{1}\approx 1-\frac{\Delta }{1-\varkappa ^{2}},\quad z_{2}\approx \Delta ,
\label{eq115}
\end{equation}%
where $\Delta =q(z_{1})=q(z_{2})\ll 1$. When approaching the border we find
(see Eqs. (\ref{eq92}) and (\ref{eq93}) when $\Delta \rightarrow 0$)%
\begin{eqnarray}
B(z,z_{1},z_{2}) &\rightarrow &\sqrt{z(1-z)}\Delta ^{-1/2}+b(z,\varkappa ), 
\notag \\
B(z,z_{2},z_{1}) &\rightarrow &-\sqrt{z(1-z)}\Delta ^{-1/2}+b(z,\varkappa ),
\notag \\
g(z,z_{1},z_{2}) &\rightarrow &v(z,\varkappa )\Delta ^{1/2},  \notag \\
\mu _{0} &\rightarrow &\frac{v(1,\varkappa )}{\pi }\Delta ^{1/2},
\label{eq116}
\end{eqnarray}%
where%
\begin{eqnarray}
b(z,\varkappa ) &=&\frac{1-(\varkappa ^{2}+2)z+\varkappa ^{2}z^{2}}{\sqrt{%
1-\varkappa ^{2}z}},  \notag \\
v(z,\varkappa ) &=&\Pi (\delta ,\varkappa ^{2},\varkappa )-2F(\delta
,\varkappa )+E(\delta ,\varkappa ).  \label{eq119}
\end{eqnarray}%
As a result, from Eqs. (\ref{eq94}), (\ref{eq95}), (\ref{eq112}), and (\ref%
{eq114}) we obtain the following solutions on the border BC:%
\begin{eqnarray}
Y_{1} &=&\chi _{1}(\tau ),  \notag \\
Y_{2} &=&\chi _{2}(\tau )+v(1,\varkappa )\chi _{1}(\tau )\xi /\pi ,
\label{eq120}
\end{eqnarray}%
where $2\pi $-periodic functions $\chi _{1,2}(\tau )$ are given by%
\begin{eqnarray}
\underset{0\leqslant \tau \leqslant \pi }{\chi _{1}(\tau )} &=&\sqrt{z(1-z)},
\notag \\
\underset{\pi \leqslant \tau \leqslant 2\pi }{\chi _{1}(\tau )} &=&-\sqrt{%
z(1-z)},  \notag \\
\underset{0\leqslant \tau \leqslant \pi }{\chi _{2}(\tau )} &=&b(z,\varkappa
)+\sqrt{z(1-z)}\left[ v(z,\varkappa )-v(1,\varkappa )\tau /\pi \right] , 
\notag \\
\underset{\pi \leqslant \tau \leqslant 2\pi }{\chi _{2}(\tau )}%
&=&b(z,\varkappa )+ \sqrt{z(1-z)}  \notag \\
&\times&\left[ v(z,\varkappa )+v(1,\varkappa )(\tau -2\pi )/\pi \right],
\label{eq121}
\end{eqnarray}%
and $\xi =\tau -\tau _{BC}$ (the constant $\tau _{B}$ is renamed to $\tau
_{BC}$ for association with the considered border).

\subsubsection{Border BD}

Considering again Eqs. (\ref{eq112}) and (\ref{eq114}) we take into account
that on the border BD%
\begin{equation*}
p^{2}=1+\frac{2\sqrt{1-\varkappa ^{2}+\varkappa ^{4}}}{1+\varkappa ^{2}}%
,\quad z_{1}=z_{2}=z_{0},
\end{equation*}%
where%
\begin{equation}
z_{0}=\frac{1}{3\varkappa ^{2}}\left( 1+\varkappa ^{2}-\sqrt{1-\varkappa
^{2}+\varkappa ^{4}}\right) .  \label{eq123}
\end{equation}%
In the vicinity of the border%
\begin{equation}
z_{1,2}\approx z_{0}\pm \Delta ^{1/2},  \label{eq124}
\end{equation}%
where%
\begin{eqnarray}
\Delta &=&\frac{1}{12\varkappa ^{4}}\left[ (3-p^{2})(1+p^{2})(1+\varkappa
^{2})^{2}-12\varkappa ^{2}\right]  \notag \\
&\approx &\frac{q_{0}-q(z_{1,2})}{\sqrt{1-\varkappa ^{2}+\varkappa ^{4}}}\ll
1,\quad q_{0}=q(z_{0}).\quad \qquad  \label{eq124a}
\end{eqnarray}%
\vspace{4pt}

When approaching the border we have (see Eqs. (\ref{eq92}) and (\ref{eq93})
when $\Delta \rightarrow 0$)%
\begin{eqnarray}
B(z,z_{1},z_{2}) &\rightarrow &\frac{z_{0}-z}{z_{0}}+\frac{2c(z,\varkappa )}{%
z_{0}}\Delta ^{1/2},  \notag \\
B(z,z_{2},z_{1}) &\rightarrow &\frac{z_{0}-z}{z_{0}}-\frac{2c(z,\varkappa )}{%
z_{0}}\Delta ^{1/2},  \notag \\
g(z,z_{1},z_{2}) &\rightarrow &\frac{2w(z,\varkappa )}{z_{0}}\Delta ^{1/2}, 
\notag \\
\mu _{0} &\rightarrow &\frac{2w(1,\varkappa )}{\pi z_{0}}\Delta ^{1/2},
\label{eq125}
\end{eqnarray}%
where%
\begin{eqnarray}
c(z,\varkappa ) &=&\frac{z}{2z_{0}}\left( \frac{1-\varkappa ^{2}z_{0}^{2}}{%
1-\varkappa ^{2}z_{0}z}+\frac{2\sqrt{q_{0}}(z_{0}-z)}{z_{0}\sqrt{q}+z\sqrt{%
q_{0}}}\right) ,  \notag \\
w(z,\varkappa ) &=&z_{0}\frac{\partial s(z,z_{0})}{\partial z_{0}}  \notag \\
&=&\varkappa ^{2}\sqrt{q_{0}}\left( \frac{\partial \Pi (\delta ,n,\varkappa )%
}{\partial n}\right) _{n=\varkappa ^{2}z_{0}}\negthickspace\negthickspace%
\negthickspace-s(z,z_{0}),\quad \qquad  \label{eq128}
\end{eqnarray}%
\begin{widetext}
\begin{equation}
\frac{\partial \Pi (\delta ,n,\varkappa )}{\partial n}=\frac{1}{2(\varkappa
^{2}-n)(n-1)}\left( E(\delta ,\varkappa )+\frac{\varkappa ^{2}-n}{n}F(\delta
,\varkappa )+\frac{n^{2}-\varkappa ^{2}}{n}\Pi (\delta ,n,\varkappa )-\frac{n%
\sqrt{1-\varkappa ^{2}\sin ^{2}\delta }\sin 2\delta }{2(1-n\sin ^{2}\delta )}%
\right) .  \label{eq129}
\end{equation}%
\end{widetext}Using these asymptotics in Eqs. (\ref{eq94}), (\ref{eq95}), (%
\ref{eq112}), and (\ref{eq114}), we obtain the solutions on the border BD:%
\begin{eqnarray}
Y_{1} &=&\chi _{1}(\tau )+w(1,\varkappa )\chi _{2}(\tau )\xi /\pi ,  \notag
\\
Y_{2} &=&\chi _{2}(\tau ),  \label{eq130}
\end{eqnarray}%
where $\xi =\tau -\tau _{BD}$, and $2\pi $-periodic functions $\chi
_{1,2}(\tau )$ are given by%
\begin{eqnarray}
\underset{0\leqslant \tau \leqslant \pi }{\chi _{1}(\tau )} &=&c(z,\varkappa
)+\frac{z_{0}-z}{z_{0}}\left[ w(z,\varkappa )-w(1,\varkappa )\tau /\pi %
\right] ,  \notag \\
\underset{\pi \leqslant \tau \leqslant 2\pi }{\chi _{1}(\tau )}
&=&-c(z,\varkappa )  \notag \\
&-&\frac{z_{0}-z}{z_{0}}\left[ w(z,\varkappa )+w(1,\varkappa )(\tau -2\pi
)/\pi \right] ,  \notag \\
\chi _{2}(\tau ) &=&\frac{z_{0}-z}{z_{0}}.\quad  \label{eq131}
\end{eqnarray}

\subsection{The Lam\'{e} equation with a slowly varying parameter}

Now we can apply our approach developed in Sec. IV to trace the evolution of
the scalar field perturbations due to cosmological expansion. As the
universe expands, the energy density $\rho $ and, hence, the parameter $%
\varkappa $ decrease slowly according to Eqs. (\ref{eq23}), (\ref{eq25}),
and (\ref{eq61}). Therefore, the points of the $(\varkappa ^{2},p^{2})$%
-plane representing the Fourier modes will move along the trajectories $%
p^{2}=p^{2}(\varkappa ,\varkappa _{0},p_{0})$ described by Eq. (\ref{eq60})
(as indicated by the arrows in Figs. 2 and 3). 
\begin{figure}[htb]
\includegraphics[width=0.48\textwidth]{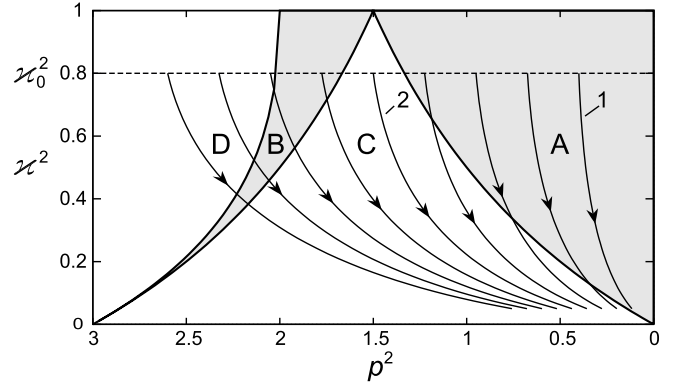}
\caption{The family of trajectories (\ref{eq60}) for $\varkappa_0^2=0.8$.}
\end{figure}
For illustration, we find
here solutions of the Lam\'{e} equation only on the trajectories with
initial points $(\varkappa _{0}^{2},p_{0}^{2})$ lying in the resonant zone A
and in the nonresonant zone C. In addition, when constructing the solutions
we restrict ourselves, for simplicity, to zero-order approximation formulas.

For trajectories lying in the zone A we find

\begin{eqnarray}
Y(\tau ,\varkappa _{0},p_{0}) &\approx &\sqrt{\frac{W_{0}(\varkappa
_{0},p_{0})}{W_{0}(\varkappa ,p)}}\bigg[A^{+}\psi _{0}(\tau ,\varkappa
,p)e^{\int_{0}^{\tau }\mu _{0}(\varkappa ,p)\,d\tau }  \notag \\
&+&A^{-}\psi _{0}(-\tau ,\varkappa ,p)e^{-\int_{0}^{\tau }\mu _{0}(\varkappa
,p)\,d\tau }\bigg]  \label{eq132}
\end{eqnarray}%
(see Eqs. (\ref{eq41}) and (\ref{eq48a})), where $\mu _{0}(\varkappa ,p)$, $%
\psi _{0}(\pm \tau ,\varkappa ,p)$ and $W_{0}(\varkappa ,p)$ are given by
Eqs. (\ref{eq87})-(\ref{eq90}), $\varkappa (\tau )$ is the solution of Eq. (%
\ref{eq25}) with (\ref{eq61}), $\varkappa _{0}=\varkappa (0)$, $p_{0}=k/m$
(we set $a(\varkappa _{0})=1$), and%
\begin{equation}
A^{\pm }(\varkappa _{0},p_{0})\approx \frac{1}{2}Y(0,\varkappa _{0},p_{0}) \mp \frac{Y_{\tau}(0,\varkappa _{0},p_{0})}{W_{0}(\varkappa _{0},p_{0})}.  \label{eq132a}
\end{equation}

Similarly, for trajectories in the zone C, we obtain

\begin{eqnarray}
Y(\tau ,\varkappa _{0},p_{0})&\approx&\sqrt{\frac{W_{0}(\varkappa _{0},p_{0})}{W_{0}(\varkappa ,p)}}\bigg[C^{+}\psi _{0}(\tau ,\varkappa,p)e^{i\int_{0}^{\tau }\nu _{0}(\varkappa ,p)\,d\tau }\notag\\
&+&C^{-}\psi _{0}(-\tau ,\varkappa ,p)e^{-i\int_{0}^{\tau }\nu_{0}(\varkappa,p)\,d\tau }\bigg],
\label{eq132b}
\end{eqnarray}%
where $\psi _{0}(\tau ,\varkappa ,p)$, $W_{0}(\varkappa ,p)$, and $\nu
_{0}(\varkappa ,p)$ are given by Eqs. (\ref{eq94})-(\ref{eq96}), and (\ref%
{eq98}) with allowance for Eq. (\ref{eq97}), and $C^+ = (C^-)^*$,%
\begin{equation}
C^{-}(\varkappa _{0},p_{0})\approx \frac{1}{2}Y(0,\varkappa _{0},p_{0}) + \frac{Y_{\tau}(0,\varkappa _{0},p_{0})}{W_{0}(\varkappa _{0},p_{0})}.  \label{eq132c}
\end{equation}

Figure 4 demonstrates excellent agreement of solutions (\ref{eq132}) and (%
\ref{eq132b}) (solid lines) with the results of direct numerical integration
of the Lam\'{e} equation (\ref{eq59}) (marked with dots) for the
trajectories labelled 1 and 2 in Fig. 3.
\begin{figure}[htb]
\includegraphics[width=0.48\textwidth]{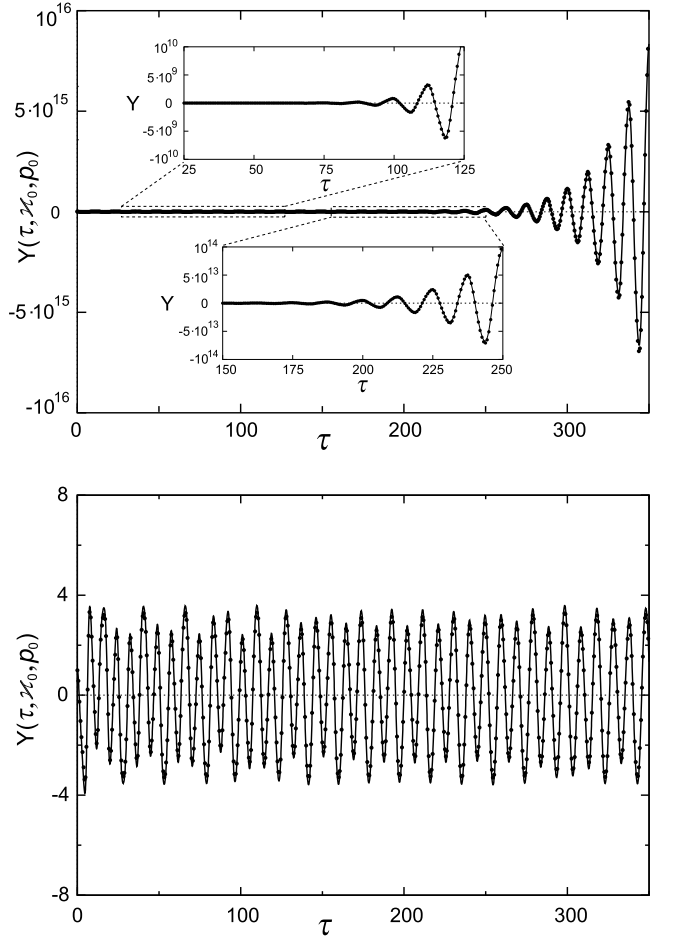}
\caption{The solutions of the Lam{\'e} equation in zone A for $p_0^2=0.4$ (top) and in zone C for $p_0^2=1.5$ (bottom). Both for $\varkappa_0^2=0.8,\quad \varepsilon=0.0025$.}
\end{figure}

Notice that formula (\ref{eq41}) loses its validity when approaching the
borders of the zones. Indeed, let the representative point, moving slowly
along the considered trajectory, cross a border at time $\tau _{B}$.
Denoting $\eta =\varepsilon (\tau -\tau _{B})$, we find that near the point
of intersection $\left\vert \mu _{0}\right\vert \sim \left\vert \eta
\right\vert ^{1/2}$, $\left\vert \mu _{1}^{\pm }\right\vert \sim \left\vert
\eta \right\vert ^{-1}$ and the solution tends to infinity as $\left\vert
\eta \right\vert ^{-1/4}$, just as it happens in the usual WKB method. In
particular, for trajectories passing through zone B this follows from Eqs. (%
\ref{eq93})-(\ref{eq96}) with allowance for asymptotics (\ref{eq124})-(\ref%
{eq125}) or (\ref{eq115}), (\ref{eq116}). Requiring $\left\vert \mu
_{0}\right\vert \gg \varepsilon \left\vert \mu _{1}^{\pm }\right\vert $, we
conclude that solution (\ref{eq41}) is valid only in the interior regions of
the zones where%
\begin{equation}
\varepsilon \left\vert \eta (\tau )\right\vert ^{-3/2}\ll 1.  \label{eq134}
\end{equation}%
The same condition follows from the requirement $\left\vert \psi
_{0}\right\vert \gg \varepsilon \left\vert \psi _{1}^{\pm }\right\vert $.

Let us examine the contribution of various Fourier modes to the field
perturbation $\delta \phi $. Assuming the spectrum of the initial
perturbations to be isotropic, we reduce expression (\ref{eq20}) to the form

\begin{eqnarray}
&&\delta \phi (t,r)=\frac{4\pi m^{2}}{a^{3/2}(\varkappa ,\varkappa _{0})\omega^{1/2}(\varkappa )}\notag\\
&&\times\frac{1}{r}\int_{0}^{\infty }Y(\tau ,\varkappa_{0},p_{0})\sin (mp_{0}r)p_{0}dp_{0},  \label{eq162}
\end{eqnarray}%
where $dp_0=m^{-1}dk.$ 
As is seen from Figs. 2 and 3, the most significant amplification of the
modes involved in Eq. (\ref{eq162}) is provided by those segments of
trajectories that lie in the interior regions of the resonant zones. On
these segments, the resonant mode amplification is mainly determined by the
factor $\exp \int_{\tau _{0}}^{\tau }\mu _{0}d\tau $, where $\tau _{0}$ is
time of entry into a resonant zone. To estimate the contribution of any
given trajectory, we consider%
\begin{equation}
{\rm {Re}}\int_{0}^{\tau }\mu _{0}d\tau =\frac{1}{\varepsilon }S(\varkappa,\varkappa _{0},p_{0}),  \label{eq163}
\end{equation}%
where%
\begin{equation}
S(\varkappa ,\varkappa _{0},p_{0})={\rm {Re}}\int_{\varkappa _{0}}^{\varkappa
}\mu _{0}(\varkappa ,p(\varkappa ,\varkappa _{0},p_{0}))\frac{d\varkappa }{%
R(\varkappa )}.  \label{eq164}
\end{equation}%
Let us fix the value of $\varkappa _{0}$, which is equivalent to specifying
the initial energy density or the initial amplitude of the background
oscillations. Integration in (\ref{eq164}) with the use of Eqs. (\ref{eq60}%
),\ (\ref{eq61}),\ (\ref{eq87}), and (\ref{eq93}) gives the surface over $%
(\varkappa ,p_{0})$-plane shown in Fig. 5. 
\begin{figure}
\includegraphics[width=0.45\textwidth]{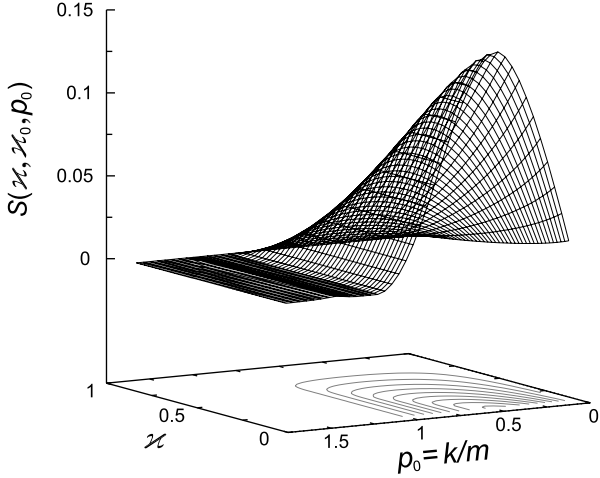}
\caption{The surface $S(\varkappa ,\varkappa _{0},p_{0})$ for $\varkappa_0^2=0.8$.}
\end{figure}
For each given $\varkappa $, the
surface has a well-defined large maximum at $p_{0}=\hat{p}_{0}(\varkappa
,\varkappa _{0})$ due to trajectories lying in zone A and a very weak
maximum due to trajectories passing through zone B. Figure 6 shows evolution
of $\hat{p}_{0}(\varkappa ,\varkappa _{0})$ and $S_{\max }(\varkappa
,\varkappa _{0})=S(\varkappa ,\varkappa _{0},\hat{p}_{0})$. With the
expansion of the universe, $\varkappa $ decreases and, consequently, $%
S_{\max }(\varkappa ,\varkappa _{0})$ increases, and $\hat{p}_{0}(\varkappa
,\varkappa _{0})$ practically does not change.

\begin{figure}[htb]
\includegraphics[width=0.40\textwidth]{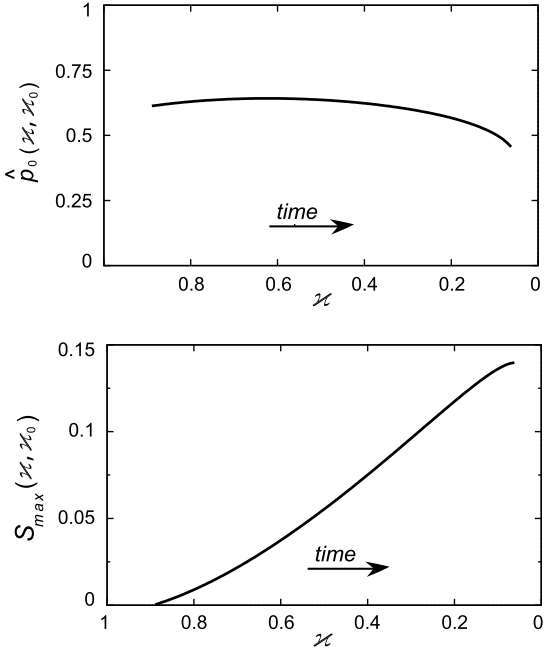}
\caption{Evolution of $\hat{p}_{0}(\varkappa ,\varkappa _{0})$ and $S_{\max }(\varkappa,\varkappa _{0})$ for $\varkappa_0^2=0.8$.}
\end{figure}

Thus, as expected, the main contribution to the integral (\ref{eq162}) is
made by the modes evolving along the trajectories lying in zone A.
Therefore, in evaluating the integral (\ref{eq162}) we can use solution (\ref%
{eq132}). Neglecting the second term in square brackets, we write it in the
form%
\begin{equation}
Y(\tau ,\varkappa _{0},p_{0})\approx \Psi (\tau ,\varkappa ,\varkappa
_{0},p_{0})e^{\frac{1}{\varepsilon }S(\varkappa ,\varkappa _{0},p_{0})},
\label{eq165}
\end{equation}%
where%
\begin{equation}
\Psi (\tau ,\varkappa ,\varkappa _{0},p_{0})=\sqrt{\frac{W_{0}(\varkappa
_{0},p_{0})}{W_{0}(\varkappa ,p)}}A^{+}(\varkappa _{0},p_{0})\psi _{0}(\tau
,\varkappa ,p),  \label{eq166}
\end{equation}%
and $p=p(\varkappa, \varkappa_0, p_0)$.
Function (\ref{eq166}) is $2\pi $-antiperiodic in $\tau $ and involves the
parameter $\varkappa $ decreasing slowly in accordance with Eqs. (\ref{eq25}%
) and (\ref{eq61}). Substituting (\ref{eq166}) into (\ref{eq162}) and taking
into account the largeness of $1/\varepsilon $, we evaluate the integral by
the Laplace method (see, e.g., \cite{Olver}). As a result, we obtain%
\begin{eqnarray}
&&\delta \phi (t,r) \approx \frac{2(2\pi )^{3/2}m^{2}\hat{p}_{0}}{%
a^{3/2}(\varkappa ,\varkappa _{0})\omega ^{1/2}(\varkappa )}\sqrt{-\frac{%
\varepsilon }{S^{\prime \prime }(\varkappa ,\varkappa _{0},\hat{p}_{0})}} \notag \\
&&\times \frac{\sin (m\hat{p}_{0}r)}{r}\Psi (\tau ,\varkappa ,\varkappa _{0},%
\hat{p}_{0})e^{\frac{1}{\varepsilon }S_{\max }(\varkappa ,\varkappa _{0})},
\label{eq167}
\end{eqnarray}%
where $a(\varkappa ,\varkappa _{0})=p_{0}/p$ (see Eq. (\ref{eq60})), $\omega
(\varkappa )$ is given in (\ref{eq56}), and $S^{\prime \prime }=\partial
^{2}S/\partial p_{0}^{2}$. Physically, this expression describes a scalar
field lump having the characteristic size $(m\hat{p}_{0})^{-1}$ and
oscillating with the background frequency $\omega $. Its amplitude grows
rapidly due to the parametric resonance phenomenon.

\section{Concluding remarks}

In this paper we have developed the asymptotic perturbation approach to
solving the Hill equation describing evolution of the perturbation Fourier
modes of the oscillating inflaton scalar field in the
Friedmann-Robertson-Walker universe. For the case of the $\phi ^{2}-\phi
^{4} $ inflaton potential we have traced evolution of the modes moving
slowly along the trajectories lying in the resonant and nonresonant zones of
the energy-momentum space. Despite the fact that we restricted ourselves to
the formulas of the lowest approximation, our approach gives solutions with
a very good accuracy even for long times of the order of $\varepsilon ^{-1}$%
. Physically, the solutions obtained give the degree of parametric
amplification when a given mode passes through the resonant zone. In order
to obtain an idea of the evolution of the entire spectrum of the
perturbations, we have analyzed the integral of the Floquet exponent along
all trajectories. We have shown that the main contribution to this integral
is provided by trajectories lying in the principal resonance zone (zone A)
and practically do not leave it. This made it possible to evaluate the
Fourier integral and find the shape and characteristic size of the
perturbation $\delta \phi $.

It should be emphasized that these results were obtained in the linear
approximation in $\delta \phi $. It is believed that at the nonlinear stage,
when $\delta \phi \gtrsim \phi $, perturbations evolve into well-localized
oscillating objects, oscillons (pulsons, by other terminology), containing a
significant part of the energy of the original condensate. For some inflaton
potentials, the formation of such soliton-like lumps was confirmed by numerical
simulation in 3+1 dimensions \cite{Amin3, Enqv, Amin4, Amin5}. Note also
that the most massive oscillons should be considered as selfgravitating
objects \cite{Kout5}, disturbing significantly the local gravitational
background. In addition, the arising oscillons can substantially change the
equation of state of the inflaton field \cite{Amin5}. These circumstances
must be taken into account both in the study of the evolution of individual
oscillons and when considering the collective effects in an ensemble of
interacting oscillons.

\section{ACKNOWLEDGEMENTS}
We are grateful to participants of the VIII-th International Conference "Solitons, collapses and turbulence" (SCT-17)  
for useful discussions.

\end{document}